\documentclass[12pt,reqno]{article}
\usepackage{amsfonts}
\usepackage{amsmath}
\usepackage{amsbsy}
\usepackage{graphicx}
\usepackage{amssymb,latexsym}
\usepackage{arydshln,leftidx}
 \numberwithin{equation}{section}

\begin{document}
 \allowdisplaybreaks[1]
\title{Cosmological Solution Moduli of Bigravity}
\author{Nejat Tevfik Y$\i$lmaz\\
Department of Electrical and Electronics Engineering,\\
Ya\c{s}ar University,\\
Sel\c{c}uk Ya\c{s}ar Kamp\"{u}s\"{u}\\
\"{U}niversite Caddesi, No:35-37,\\
A\u{g}a\c{c}l\i Yol, 35100,\\
Bornova, \.{I}zmir, Turkey.\\
\texttt{nejat.yilmaz@yasar.edu.tr}} \maketitle
\begin{abstract}
We construct the complete set of metric-configuration solutions of
the ghost-free massive bigravity for the scenario in which the
$g-$metric is the Friedmann-Lemaitre-Robertson-Walker (FLRW) one,
and the interaction Lagrangian between the two metrics contributes
an effective ideal fluid energy-momentum tensor to the g-metric
equations. This set corresponds to the exact background
cosmological solution space of the theory.
\end{abstract}

\section{Introduction}
The massive gravity theory which was constructed in
\cite{dgrt1,dgrt2} as a two-parameter family of actions, is a
nonlinear generalization of the Fierz- Pauli massive gravity model
\cite{fp}. In \cite{dgrt2} it was shown that one of the actions of
this theory is Boulware-Deser (BD)-ghost-free \cite{BD1,BD2} up to
the fourth order in metric perturbations around flat space. An
extension of this massive gravity model which originally admits a
flat reference metric was also constructed in \cite{hr1,hr2,hr3}
for a general background or reference metric. In \cite{hr2}, it
was shown that the two-parameter family of actions given in
\cite{dgrt2} are BD-ghost-free at the complete nonlinear level for
all perturbation orders. In \cite{hr3}, the analysis is extended
to show that the general reference metric massive gravity theory
in \cite{hr1} is also BD-ghost-free at the complete nonlinear
level for all orders. Later on, a ghost-free massive bimetric
theory in which the interaction term between the foreground, and
the background metrics arises from the mass terms was proposed by
introducing a copy of general relativity (GR) dynamics for the
background metric \cite{hrbg}. Within this theory, a particular
class of cosmological solutions of the coupled field equations of
the two-metric-sectors \cite{hrbg,bac1,bac2,bac3} have been
extensively studied in recent years
\cite{bac3,com1,com2,bm2,bm1,3,4,6,8,9,10,11,12,yeni1,yeni2}. The
solution space corresponding to an effective decoupling of the two
metric sectors of the theory is constructed in \cite{decoup}. The
elements of the decoupling solution moduli that is derived in
\cite{decoup} give rise to self-accelerating cosmologies in the
$g-$sector via the contribution of an effective cosmological
constant. On the other hand, in this work we will construct the
general cosmological solution space of the massive bigravity
theory. Throughout our derivation, we will follow a parallel route
with those of \cite{massgrav,mgrphysfluid} in which cosmological
solutions of massive gravity are derived in a formal fashion
without explicit classification. On the contrary, in what follows
likewise in \cite{decoup}, our main perspective will be to derive
the entire set of metric couples $(f,g)$ explicitly, which will
consistently solve all the field equations of the theory, and
which will enable a semi-decoupling of the $f-$metric from the
$g-$metric sector. Since we will consider a homogeneous, and an
isotropic scenario in the $g-$sector apart from assigning a
Friedmann-Lemaitre-Robertson-Walker (FLRW) form for $g$  which
corresponds to the cosmological background metric of the universe,
we will also take the overall contribution of the interaction
Lagrangian to the $g-$metric equations as an effective ideal fluid
energy-momentum tensor. In this respect, the Bianchi identity of
the interaction terms in the $g-$metric equations will naturally
transform to be the continuity equation of the effective fluid. We
will show that when one proposes such a solution ansatz then the
only remaining task to derive the set of metric couples which
allow this picture is to find the general solutions of an
inhomogeneous cubic matrix equation. We will derive the entire set
of solutions of this matrix equation whose coefficients are
functions of the elementary symmetric polynomials of the solutions
themselves rather than being constants. By using the general
solutions of this matrix equation one can construct the solution
moduli of the metrics $(f,g)$ which admit a cosmological scenario
in the $g-$sector.

In Section one, starting from the bigravity dynamics by assuming a
cosmological $g-$sector and thus, introducing the above-mentioned
ansatz for the contribution of the interaction terms in the
$g-$metric equations we will derive an algebraic matrix equation
whose solutions will lead us to the metric couples which are
compatible with the field equations of the cosmological picture.
In Section two, we will derive the Jordan normal form solutions of
this equation. By using this complete set of Jordan normal form
solutions, and other three sets constructed from them we will
discuss in Section three that when special form of similarity
transformations are used one can obtain the general set of
solutions of the above-mentioned ansatz-generated matrix equation.
An outline of the proof of the completeness of this set of
solutions will be given in the Appendix. Then, again in Section
three as a consequence of the completely-derived general solution
space of the ansatz equation we will be able to define the
cosmological solution moduli space of bigravity. We reserve
Section four for the derivation of the equations of the
cosmological dynamics in $g-$sector, as well as a discussion about
the associated $f-$dynamics, and its solution methodology. We will
also present the outline of the $f$-sector solution construction
of an example.
\section{The set-up}
The ghost-free bigravity action can be given as
\cite{hrbg,bac1,bac2,bac3}
\begin{subequations}\label{e1}
\begin{align}
 S=&-\frac{1}{16\pi G}\int dx^4\sqrt{-g}\bigg[ R^g+\Lambda^g -2m^2\mathcal{L}_{int}(\sqrt{\Sigma})\bigg]+S_{M}^g\notag\\
 &-\frac{\kappa}{16\pi G}\int dx^4\sqrt{-f}\bigg[ R^f+\Lambda^f \bigg]+\epsilon
 S_{M}^f,\tag{\ref{e1}}
\end{align}
\end{subequations}
 where $g$ is the foreground, and $f$ is the
background metric which are coupled to two types of matter via the
actions $S_{M}^g,S_{M}^f$, respectively. $\Lambda^g,\Lambda^f$ are
the cosmological constants in each sector. $R^g,R^f$ are the
corresponding Ricci scalars. The Lagrangian which describes the
interaction between the two metrics above reads
\begin{equation}\label{e2}
\mathcal{L}_{int}(\sqrt{\Sigma})=\beta_{1}
e_{1}(\sqrt{\Sigma})+\beta_{2} e_{2}(\sqrt{\Sigma})+\beta_{3}
e_{3}(\sqrt{\Sigma}),
\end{equation}
where $\{e_n\}$ are the elementary symmetric polynomials
\begin{subequations}\label{e3}
\begin{align}
e_{1}\equiv e_{1}(\sqrt{\Sigma})&=tr\sqrt{\Sigma},\notag\\
e_{2}\equiv e_{2}(\sqrt{\Sigma})&=\frac{1}{2}\big((tr\sqrt{\Sigma})^2-tr(\sqrt{\Sigma})^2\big),\notag\\
e_{3}\equiv
e_{3}(\sqrt{\Sigma})&=\frac{1}{6}\big((tr\sqrt{\Sigma})^3-3\,tr\sqrt{\Sigma}\:tr(\sqrt{\Sigma})^2+2\,tr(\sqrt{\Sigma})^3\big),
\tag{\ref{e3}}
\end{align}
\end{subequations}
corresponding to the square-root-matrix
\begin{equation}\label{e3.1}
\sqrt{\Sigma}=\sqrt{g^{-1}f}.
\end{equation}
Originally, the interaction Lagrangian also contains the terms
$\beta_0 e_0=\beta_0$, and $\beta_4
e_4=\beta_4\text{det}\sqrt{\Sigma}$. However, we combine their
contributions with the cosmological constants $\Lambda_g$, and
$\Lambda_f$, respectively. Eq. \eqref{e2} reduces to the
Fierz-Pauli form in the weak-field limit when one chooses
\cite{hrbg}
\begin{equation}\label{e3.2}
\beta_1+2\beta_2+\beta_3=-1.
\end{equation}
From Eq. \eqref{e1} one can obtain the field equations for the
metric $g$ as
\begin{equation}\label{e4}
R^g_{\mu\nu}-\frac{1}{2}R^g g_{\mu\nu}-\frac{1}{2}\Lambda^g
g_{\mu\nu}+m^2\mathcal{T}^g_{\mu\nu}=8\pi GT^{g}_{M\,\mu\nu}.
\end{equation}
Whereas, the field equations of the metric $f$ become
\begin{equation}\label{e4.1}
\kappa\big[R^f_{\mu\nu}-\frac{1}{2}R^f
f_{\mu\nu}-\frac{1}{2}\Lambda^f
f_{\mu\nu}\big]+m^2\mathcal{T}^f_{\mu\nu}=\epsilon 8\pi
GT^{f}_{M\,\mu\nu}.
\end{equation}
The corresponding energy-momentum tensors arising from the
interaction term Eq.\eqref{e2} are
\begin{equation}\label{e5}
\mathcal{T}^g_{\mu\nu}=-g_{\mu\rho}\tau^{\rho}_{\:\:\nu}+\mathcal{L}_{int}g_{\mu\nu},
\end{equation}
and
\begin{equation}\label{e6}
\mathcal{T}^f_{\mu\nu}=\frac{\sqrt{-g}}{\sqrt{-f}}f_{\mu\rho}\tau^{\rho}_{\:\:\nu}.
\end{equation}
The matrix $\tau$ with the entries $\{\tau^{\rho}_{\:\:\nu}\}$ is
defined to be \cite{bac3}
\begin{equation}\label{e7}
\tau=\beta_3(\sqrt{\Sigma})^3-(\beta_2+\beta_3
e_1)(\sqrt{\Sigma})^2+(\beta_1+\beta_2 e_1+\beta_3
e_2)\sqrt{\Sigma}.
\end{equation}
The effective energy-momentum tensors in Eqs. \eqref{e4}, and
\eqref{e4.1} are ought to satisfy the Bianchi identities
\begin{equation}\label{e7.6}
\nabla^g_\mu(\mathcal{T}^{g})^{\mu}_{\:\:\nu}=0,\quad
\nabla^f_\mu(\mathcal{T}^{f})^{\mu}_{\:\:\nu}=0.
\end{equation}
If a solution configuration satisfies one of these equations then
the other one is automatically satisfied \cite{bm2,bm1}. Now, let
us focus on the cosmological solutions in the $g-$sector of the
action Eq. \eqref{e1}. Thus, we will take $g$ as the FLRW metric
\begin{equation}\label{e7.7}
g=-dt^2+\frac{a^2(t)}{1-kr^2}dr^2+a^2(t)r^2 d\theta^2 +a^2(t)r^2
sin^2 \theta d\varphi^2.
\end{equation}
Let us also consider the solutions for which the effective
energy-momentum tensor entering into the $g-$metric equations in
Eq. \eqref{e4} takes the form
\begin{equation}\label{e8}
\mathcal{T}^g_{\mu\nu}=(\tilde{\rho}(t)+\tilde{p}(t))U_{\mu}U_{\nu}+\tilde{p}(t)g_{\mu\nu},
\end{equation}
of an ideal fluid. For an ideal fluid the on-shell Lagrangian can
be taken as \cite{mgrphysfluid}
\begin{equation}\label{e8.5}
\mathcal{L}_{IF}=\tilde{p},
\end{equation}
so that
\begin{equation}\label{e8.55}
\mathcal{T}^{g}_{\mu\nu}=-2\frac{1}{\sqrt{-g}}\frac{\delta(\sqrt{-g}\mathcal{L}_{IF})}{\delta(g^{\mu\nu})}.
\end{equation}
Therefore, for the solutions which generate Eq. \eqref{e8} in the
field equations, Eq. \eqref{e4} a close inspection of the
interaction term in the action Eq. \eqref{e1}, and Eq.
\eqref{e8.55} shows us that we must have
\begin{equation}\label{e8.6}
\mathcal{L}_{int}=\tilde{p}.
\end{equation}
This is due to the fact that while Eq. \eqref{e5} is derived by
varying the interaction term in the action Eq. \eqref{e1} with
respect to the metric $g$, Eq. \eqref{e8.55} is obtained by
varying the Lagrangian Eq. \eqref{e8.5} with respect to $g$, and
by using the first law of thermodynamics \cite{mgrphysfluid}. Our
effective fluid that is introduced in Eq. \eqref{e8} will
certainly obey the first law of thermodynamics as, it must satisfy
the conservation equation in Eq. \eqref{e7.6} which will result in
an ordinary continuity or fluid equation when Eq. \eqref{e8} is
substituted in it. If we take the effective ideal fluid
four-velocity vector as $U_{\mu}=(1,0,0,0)$ in the rest frame of
the fluid, and use the FLRW metric Eq. \eqref{e7.7} we obtain
\begin{equation}\label{e9}
g^{-1}\mathcal{T}^g= \left(\begin{matrix}
-\tilde{\rho}&\texttt{0}&\texttt{0}&\texttt{0}
\\
\texttt{0}&\tilde{p}&\texttt{0}&\texttt{0}\\\texttt{0}&\texttt{0}&\tilde{p}
&\texttt{0}\\\texttt{0}&\texttt{0}
&\texttt{0}&\tilde{p}\end{matrix}\right).
\end{equation}
Index raising on both sides of Eq. \eqref{e5} by the metric $g$
gives
\begin{equation}\label{e9.5}
(\mathcal{T}^{g})^{\:\mu}_{\:\:\nu}=-\tau^{\mu}_{\:\:\nu}+\mathcal{L}_{int}\delta^{\mu}_{\:\:\nu},
\end{equation}
where
$(\mathcal{T}^{g})^{\:\mu}_{\:\:\nu}=[g^{-1}\mathcal{T}^g]^{\mu}_{\:\:\nu}$.
By using Eqs. \eqref{e8.6}, and \eqref{e9} in this expression we
obtain the matrix equation
\begin{equation}\label{e10}
A(\sqrt{\Sigma})^3+B(\sqrt{\Sigma})^2+C\sqrt{\Sigma}+\mathcal{D}=0,
\end{equation}
with
\begin{equation}\label{e10.5}
\mathcal{D}= \left(\begin{matrix}
D&\texttt{0}&\texttt{0}&\texttt{0}
\\
\texttt{0}&\texttt{0}&\texttt{0}&\texttt{0}\\\texttt{0}&\texttt{0}&\texttt{0}
&\texttt{0}\\\texttt{0}&\texttt{0}
&\texttt{0}&\texttt{0}\end{matrix}\right),
\end{equation}
where
\begin{subequations}\label{e11}
\begin{align}
A&=\beta_3,\notag\\
B&=-\beta_2-\beta_3e_1,\notag\\
C&=\beta_1+\beta_2e_1+\beta_3e_2,\notag\\
D&=-\tilde{\rho}-\tilde{p}.\tag{\ref{e11}}
\end{align}
\end{subequations}
\section{Classification of the solutions}
Next, we will derive and classify the Jordan canonical form
solutions of the cubic matrix equation \eqref{e10}. This is a
highly non-trivial matrix equation for two reasons: first, it is
not in a polynomial form, and second, its coefficients are
functions of the elementary symmetric polynomials $e_1,e_2$ of its
solutions $\sqrt{\Sigma}$ rather than being constants. For this
reason, in this section we will derive the diagonal and the
nondiagonal Jordan form solutions of it, and show in the next
section that they can be used to generate the entire solution
space. Firstly, let us define the polynomials
\begin{equation}\label{e12}
Ax^3+Bx^2+Cx+D=0,
\end{equation}
and
\begin{equation}\label{e12.1}
x(Ax^2+Bx+C)=0,
\end{equation}
whose roots we will generally call $\alpha_i$, and $\lambda_j$,
respectively. Since, in general for any Jordan canonical form
matrix $J$ (diagonal or nondiagonal) when it is substituted into
the Eq. \eqref{e10} the eigenvalues namely the diagonal elements
of $J$ must satisfy one copy of Eq. \eqref{e12}, and three copies
of the polynomial in Eq. \eqref{e12.1} the multiplicity of
$\alpha_i$ in the diagonal of $J$ must be one. This is obvious for
the diagonal Jordan forms. Besides, if we have a nondiagonal
Jordan canonical form as
\begin{subequations}\label{e12.11}
\begin{align}
J=\left(\begin{matrix} \alpha_i&\texttt{1}&\texttt{0}&\texttt{0}
\\
\texttt{0}&\alpha_i&\bullet&\texttt{0}\\\texttt{0}&\texttt{0}&\bullet
&\bullet\\\texttt{0}&\texttt{0}
&\texttt{0}&\bullet\end{matrix}\right),
 \tag{\ref{e12.11}}
\end{align}
\end{subequations}
when this solution ansatz is used in Eq. \eqref{e10} then the
diagonal entries would lead to two inconsistent equations
$A(\alpha_i)^3+B(\alpha_i)^2+C(\alpha_i)+D=0$, and
$A(\alpha_i)^3+B(\alpha_i)^2+C(\alpha_i)=0$\footnote{Obviously,
these two equations can be satisfied simultaneously if $D=0$ which
would correspond to $\tau=0$ cases. However, we take $D\neq 0$ so
that the effective fluid is not simply an effective cosmological
constant. Thus, we exclude the $\tau=0$ solutions here as they are
completely derived in \cite{decoup}.}. Therefore, following this
observation we can conclude that the Jordan form solutions of Eq.
\eqref{e10} are partitioned as
\begin{equation}\label{e13}
J={\left(\begin{array}{c|c}  \alpha_i &
0    \\
\hline 0& H_{3\times 3}
\end{array}\right)},
\end{equation}
where $H_{3\times 3}$ is a three by three Jordan normal form
matrix which satisfies the matrix polynomial equation
\begin{equation}\label{e14}
A(H_{3\times 3})^3+B(H_{3\times 3})^2+CH_{3\times 3}=0.
\end{equation}
The diagonal ones are naturally in the form of Eq. \eqref{e13},
and the nondiagonal ones must be in this form due to the
partitioning nogo fact discussed above. $H_{3\times 3}$ must be
constructed by its eigenvalues $\{\lambda_j\}$. As a result, we
deduce that the classification of the Jordan form solutions of Eq.
\eqref{e10} must be based on the classification of the solutions
of Eq. \eqref{e14}. We should remark one important point here
that, in deriving these solutions we will exclude the cases which
arise from the conditions $B=0$, and/or $C=0$. As it will be clear
in Section four, such conditions would lead to extra constraints
on the equation of state $\tilde{p}=\tilde{p}(\tilde{\rho})$ of
the effective fluid that would cause it to be nondynamical. In
addition, we will also exclude the trivial case of $H_{3\times
3}=0$.
\subsection{$\Delta>0$ Solutions}
Let us first consider the solutions when $\Delta=B^2-4AC>0$. In
this case there are three distinct roots of the polynomial in Eq.
\eqref{e12.1}, they are $\{0,\lambda_1,\lambda_2\}$. Now, none of
the Jordan normal forms which satisfy Eq. \eqref{e14} may have a
repeated root of its minimum polynomial which is formed by a
subset of the factors in Eq. \eqref{e14}. That is to say, for
these cases if we write Eq. \eqref{e14} in the form
\begin{equation}\label{e14.1}
H(H-\lambda_1\mathbf{1}_3)(H-\lambda_2\mathbf{1}_3)=0,
\end{equation}
where $\mathbf{1}_3$ is the unit $3\times 3$ matrix then we see
that the solutions of this equation must make the product of three
factors, or any two factors, or just a single factor vanish. Thus,
there are $1+3+3$ distinct classes of solutions where one of them
is trivial. For each class the vanishing combination of factors
become the minimum polynomial of the corresponding $3\times 3$
matrix solution. Therefore, we also observe that for each of these
classes the roots of the corresponding minimum polynomial are not
repeated (as $\lambda_1,\lambda_2$ are distinct)\footnote{We
should note that, if $\lambda_{1}$, or $\lambda_2$ vanish then
there will be a repeated root (the zero root). In this case a
single nondiagonalizable Jordan form exits but this case requires
$C=0$ condition that we exclude as we pointed above.}. Since a
matrix is diagonalizable if and only if its minimum polynomial has
no repeated roots we deduce that in this case any solution
satisfying Eq. \eqref{e14} must be a diagonalizable one. On the
other hand, if a Jordan canonical form satisfies Eq. \eqref{e14}
its similarity equivalence class also does. Hence, combining these
two facts we conclude that all the Jordan forms which satisfy Eq.
\eqref{e14} must be diagonal. The reader may also verify this
result by direct substitution. In other words, none of the
nondiagonal Jordan canonical forms whose diagonal elements are
chosen from the set $\{0,\lambda_1,\lambda_2\}$ satisfies Eq.
\eqref{e14}. Therefore, upon this identification of the $3 \times
3$ sectors, the corresponding $4 \times 4$ Jordan normal forms
which satisfy Eq. \eqref{e10} can be listed as
\begin{subequations}\label{e15}
\begin{align}
 N_1&=\left(\begin{matrix}
\alpha_{i}&\texttt{0}&\texttt{0}&\texttt{0}
\\
\texttt{0}&\lambda_1&\texttt{0}&\texttt{0}\\\texttt{0}&\texttt{0}&\lambda_1
&\texttt{0}\\\texttt{0}&\texttt{0}
&\texttt{0}&\lambda_1\end{matrix}\right), N_2=\left(\begin{matrix}
\alpha_i&\texttt{0}&\texttt{0}&\texttt{0}
\\
\texttt{0}&\lambda_2&\texttt{0}&\texttt{0}\\\texttt{0}&\texttt{0}&\lambda_2
&\texttt{0}\\\texttt{0}&\texttt{0}
&\texttt{0}&\lambda_2\end{matrix}\right), N_3=
\left(\begin{matrix} \alpha_i&\texttt{0}&\texttt{0}&\texttt{0}
\\
\texttt{0}&\texttt{0}&\texttt{0}&\texttt{0}\\\texttt{0}&\texttt{0}&\lambda_1
&\texttt{0}\\\texttt{0}&\texttt{0}
&\texttt{0}&\lambda_1\end{matrix}\right),\notag\\\notag\\
N_4&=\left(\begin{matrix}
\alpha_{i}&\texttt{0}&\texttt{0}&\texttt{0}
\\
\texttt{0}&\texttt{0}&\texttt{0}&\texttt{0}\\\texttt{0}&\texttt{0}&\texttt{0}
&\texttt{0}\\\texttt{0}&\texttt{0}
&\texttt{0}&\lambda_1\end{matrix}\right), \quad
N_5=\left(\begin{matrix} \alpha_i&\texttt{0}&\texttt{0}&\texttt{0}
\\
\texttt{0}&\texttt{0}&\texttt{0}&\texttt{0}\\\texttt{0}&\texttt{0}&\lambda_2
&\texttt{0}\\\texttt{0}&\texttt{0}
&\texttt{0}&\lambda_2\end{matrix}\right), \quad N_6=
\left(\begin{matrix} \alpha_i&\texttt{0}&\texttt{0}&\texttt{0}
\\
\texttt{0}&\texttt{0}&\texttt{0}&\texttt{0}\\\texttt{0}&\texttt{0}&\texttt{0}
&\texttt{0}\\\texttt{0}&\texttt{0}
&\texttt{0}&\lambda_2\end{matrix}\right),
\quad\quad\quad\quad\quad\quad\tag{\ref{e15}}
\end{align}
\end{subequations}
as well as the ones,
\begin{subequations}\label{e16}
\begin{align}
N_7=\left(\begin{matrix}
\alpha_{i}&\texttt{0}&\texttt{0}&\texttt{0}
\\
\texttt{0}&\lambda_1&\texttt{0}&\texttt{0}\\\texttt{0}&\texttt{0}&\lambda_1
&\texttt{0}\\\texttt{0}&\texttt{0}
&\texttt{0}&\lambda_2\end{matrix}\right), N_8=\left(\begin{matrix}
\alpha_i&\texttt{0}&\texttt{0}&\texttt{0}
\\
\texttt{0}&\lambda_2&\texttt{0}&\texttt{0}\\\texttt{0}&\texttt{0}&\lambda_2
&\texttt{0}\\\texttt{0}&\texttt{0}
&\texttt{0}&\lambda_1\end{matrix}\right),\notag\\\notag\\ N_9=
\left(\begin{matrix} \alpha_i&\texttt{0}&\texttt{0}&\texttt{0}
\\
\texttt{0}&\lambda_1&\texttt{0}&\texttt{0}\\\texttt{0}&\texttt{0}&\lambda_2
&\texttt{0}\\\texttt{0}&\texttt{0}
&\texttt{0}&\texttt{0}\end{matrix}\right),
\quad\quad\quad\quad\quad\quad\tag{\ref{e16}}
\end{align}
\end{subequations}
in which both of the roots $\lambda_1,\lambda_2$ appear. By direct
substitution, the reader may verify that these matrices do solve
Eq. \eqref{e10}, and the corresponding nondiagonal Jordan forms
that share the same eigenvalues can not satisfy Eq. \eqref{e10} in
this case when $\Delta>0$. To find the explicit form of these
solutions we have to know $e_1,e_2$ which constitute both the
coefficients given in Eq. \eqref{e11} (of the matrix equation that
these solutions must satisfy), and the entries of these solution
matrices listed above. In other words, we have to solve $e_1$, and
$e_2$ in terms of the $\{\beta_i\}-$parameters of the action Eq.
\eqref{e1}, and the constituents of the solution ansatz Eq.
\eqref{e8} so that Eq. \eqref{e10} is satisfied. We will first
consider the cases $N_{1,2,3,4,5,6}$. If we take the trace of
these solutions we get
\begin{equation}\label{e17}
e_1=tr(N_a)=\alpha_i+n\lambda_j,
\end{equation}
where $n=3,3,2,1,2,1$ for $N_1,N_2,N_3,N_4,N_5,N_6,$ respectively.
We also have
\begin{equation}\label{e18}
tr(N_a)^2=(e_1)^2-2e_2=(\alpha_i)^2+n(\lambda_j)^2.
\end{equation}
By using Eq. \eqref{e17}, and singling out $e_2$ from this
expression we get
\begin{equation}\label{e18.5}
e_2=-\frac{n+n^2}{2}(\lambda_j)^2+ne_1\lambda_j.
\end{equation}
If we substitute this result into Eq. \eqref{e12.1} we obtain the
relation
\begin{equation}\label{e19}
a(\lambda_j)^2+b\lambda_j+c=0,
\end{equation}
where
\begin{equation}\label{e20}
a=\frac{(2-n-n^2)}{2}\beta_3,\quad b=(n-1)\beta_3
e_1-\beta_2,\quad c=\beta_1+\beta_2 e_1.
\end{equation}
Since from Eq. \eqref{e17} we have $\alpha_i=e_1-n\lambda_j$,
substituting this into Eq. \eqref{e12}, and successive usage of
Eq. \eqref{e19} leads us to the relation
\begin{subequations}\label{e21}
\begin{align}
\lambda_j=&\bigg[\frac{(n^2-n)\beta_3}{2}\bigg(-\frac{b}{a}e_1-n\frac{b^2}{a^2}+n\frac{c}{a}\bigg)+\beta_2ne_1+\beta_2n^2\frac{b}{a}-\beta_1n\bigg]^{-1}
\notag\\
\quad &\times\bigg[\frac{(n^2-n)\beta_3
ce_1}{2a}+\frac{n(n^2-n)\beta_3bc}{2a^2}-\frac{n^2c\beta_2}{a}-\beta_1e_1-D\bigg].\tag{\ref{e21}}
\end{align}
\end{subequations}
Finally, when we use Eqs. \eqref{e21}, and \eqref{e18.5} back in
Eq. \eqref{e19} and we refer to the definitions in Eq. \eqref{e20}
we obtain an equation for $e_1$ solely in terms of the
$\beta_i-$coefficients. For the $n=3$ cases this equation reads
\begin{equation}\label{e22}
a_3(e_1)^4+b_3(e_1)^3+c_3(e_1)^2+d_3e_1+f_3=0,
\end{equation}
where we define
\begin{subequations}\label{e23}
\begin{align}
a_3&=3(\beta_3)^2(-3(\beta_2)^2+4\beta_1\beta_3),\notag\\
b_3&=6\beta_3(-15(\beta_2)^3+20\beta_1\beta_2\beta_3+2(\beta_3)^2D),\notag\\
c_3&=-216(\beta_2)^4+159\beta_1(\beta_2)^2\beta_3+172(\beta_1)^2(\beta_3)^2+102\beta_2(\beta_3)^2D,\notag\\
d_3&=204\beta_1\beta_2(-3(\beta_2)^2+4\beta_1\beta_3)+\beta_3(249(\beta_2)^2+20\beta_1\beta_3)D,\notag\\
f_3&=-432(\beta_1)^2(\beta_2)^2+576(\beta_1)^3\beta_3+36(\beta_2)^3D+240\beta_1\beta_2\beta_3D-125(\beta_3)^2D^2.\tag{\ref{e23}}
\end{align}
\end{subequations}
For the $n=2$ cases we get
\begin{equation}\label{e24}
a_2(e_1)^3+b_2(e_1)^2+c_2e_1+d_2=0,
\end{equation}
where
\begin{subequations}\label{e25}
\begin{align}
a_2&=-(\beta_2)^3\beta_3+\beta_1\beta_2(\beta_3)^2,\notag\\
b_2&=-6(\beta_2)^4+4\beta_1(\beta_2)^2\beta_3+2(\beta_1)^2(\beta_3)^2+\beta_2(\beta_3)^2D,\notag\\
c_2&=-21\beta_1(\beta_2)^3+21(\beta_1)^2\beta_2\beta_3+8(\beta_2)^2\beta_3D-2\beta_1(\beta_3)^2D,\notag\\
d_2&=-18(\beta_1)^2(\beta_2)^2+18(\beta_1)^3\beta_3+3(\beta_2)^3D+6\beta_1\beta_2\beta_3D-4(\beta_3)^2D^2.
\tag{\ref{e25}}
\end{align}
\end{subequations}
For each real root of Eq. \eqref{e22} we have the solutions
$N_1,N_2$, and for each real root of Eq. \eqref{e24} we have the
solutions $N_3,N_5$ of Eq. \eqref{e10} with the corresponding
entries that can be read from
\begin{equation}\label{e26}
\lambda_j=\frac{-b\pm\sqrt{b^2-4ac}}{2a},\quad
\alpha_i=e_1-n\lambda_j.
\end{equation}
The domain of validity of these solutions are determined by the
conditions $B^2-4AC>0$, and $b^2-4ac\geq0$, together with the
reality conditions of the corresponding roots of the Eqs.
\eqref{e22}, and \eqref{e24} which can be obtained from the
definitions of the coefficients in Eqs. \eqref{e23}, and
\eqref{e25}. These conditions will define a validity domain for
each particular solution in the union of the
$\{\beta_i\}-$parameter space, and the state space of the
effective ideal fluid. On the other hand, for the cases with
$n=1$, namely for the solutions of the form $N_4,N_6$ we have a
simpler picture. In these cases, Eq. \eqref{e19} gives
\begin{equation}\label{e27}
\lambda_j=e_1+\frac{\beta_1}{\beta_2},
\end{equation}
and from Eq. \eqref{e18.5} we have
\begin{equation}\label{e28}
e_2=-(e_1+\frac{\beta_1}{\beta_2})^2+e_1(e_1+\frac{\beta_1}{\beta_2}).
\end{equation}
Now, by using Eq. \eqref{e27} in Eq. \eqref{e17} we get
\begin{equation}\label{e29}
\alpha_i=-\frac{\beta_1}{\beta_2}.
\end{equation}
Substituting this result, together with Eq. \eqref{e28} into Eq.
\eqref{e12} gives us
\begin{equation}\label{e30}
e_1=-2\frac{\beta_1}{\beta_2}+\frac{D}{\beta_1}.
\end{equation}
The Eqs. \eqref{e27}, \eqref{e29}, and \eqref{e30} define the
explicit form of the entries of $N_4,N_6$ in terms of the
parameters of the theory, and the state of the effective fluid.
The validity domain of these solutions is defined from the
condition $B^2-4AC>0$ without extra requirements. Now, let us
consider the solutions $N_7,N_8$. Taking the trace of these
solutions we get
\begin{equation}\label{e31}
e_1=trN_a=\alpha_i+2\lambda_j+\lambda_k
=\alpha_i+\lambda_j-\frac{B}{A},
\end{equation}
where we label the excess or the repeated root on the diagonal of
the solution by $\lambda_j$. By refereing to the definitions in
Eq. \eqref{e11} we find that
\begin{equation}\label{e32}
\alpha_i=-\lambda_j-\frac{\beta_2}{\beta_3}.
\end{equation}
We also have
\begin{equation}\label{e33}
tr(N_a)^2=(e_1)^2-2e_2=(\alpha_i)^2+2(\lambda_j)^2+(\lambda_k)^2.
\end{equation}
By using the identity
\begin{equation}\label{e34}
(\lambda_j)^2+(\lambda_k)^2 =\frac{B^2}{A^2}-\frac{2C}{A},
\end{equation}
in the above equation we see that for these solutions
\begin{equation}\label{e35}
\alpha_i\lambda_j=\frac{(\beta_2)^2}{(\beta_3)^2}-\frac{\beta_1}{\beta_3}.
\end{equation}
By using this result in Eq. \eqref{e32} we obtain the equation
\begin{equation}\label{e36}
-(\lambda_j)^2-\frac{\beta_2}{\beta_3}\lambda_j+\frac{\beta_1}{\beta_3}-\frac{(\beta_2)^2}{(\beta_3)^2}=0,
\end{equation}
for $\lambda_j$. Its solutions are
\begin{equation}\label{e37}
\lambda^{\pm}_j=-\frac{1}{2}\bigg(\frac{\beta_2}{\beta_3}\pm\sqrt{-3\frac{(\beta_2)^2}{(\beta_3)^2}+4\frac{\beta_1}{\beta_3}}\:\bigg).
\end{equation}
Using Eqs. \eqref{e36}, and \eqref{e37} in Eq. \eqref{e12.1} will
enable us to write $e_2$ in terms of $e_1$. After some algebra we
get
\begin{equation}\label{e38}
e_2=\bigg(-\frac{3}{2}\frac{\beta_2}{\beta_3}\mp\frac{1}{2}\sqrt{-3\frac{(\beta_2)^2}
{(\beta_3)^2}+4\frac{\beta_1}{\beta_3}}\bigg)e_1\mp\frac{\beta_2}{\beta_3}\sqrt{-3\frac{(\beta_2)^2}
{(\beta_3)^2}+4\frac{\beta_1}{\beta_3}}-2\frac{\beta_1}{\beta_3}.
\end{equation}
We note that, when the $(+)$ solution is taken in Eq. \eqref{e37}
the opposite sign must be chosen in Eq. \eqref{e38}, and vice
versa. Finally, substituting expressions \eqref{e32}, \eqref{e37},
and \eqref{e38} into Eq. \eqref{e12} will lead us to the explicit
value of $e_1$ for either of the solutions in Eq. \eqref{e37}. For
the solutions $\lambda^\pm_j$ this computation reads
\begin{equation}\label{e39}
e_1^\pm=-\frac{2\beta_2}{\beta_3} -\frac{2(\beta_3)^2D}
{\tilde{C^\pm}},
\end{equation}
where we defined
\begin{equation}\label{e39.5}
\tilde{C^\pm}=3(\beta_2)^2\beta_3-4\beta_1(\beta_3)^2\pm\beta_2(\beta_3)^2\sqrt{-3\frac{(\beta_2)^2}
{(\beta_3)^2}+4\frac{\beta_1}{\beta_3}}.
\end{equation}
In summary, for these latest cases we find that
\begin{subequations}\label{e40}
\begin{align}
\quad\quad\quad N_7=\left(\begin{matrix}
\alpha^+_{i}&\texttt{0}&\texttt{0}&\texttt{0}
\\
\texttt{0}&\lambda^+_j&\texttt{0}&\texttt{0}\\\texttt{0}&\texttt{0}&\lambda^+_j
&\texttt{0}\\\texttt{0}&\texttt{0}
&\texttt{0}&\lambda^+_k\end{matrix}\right),\quad
N_8=\left(\begin{matrix}
\alpha^-_i&\texttt{0}&\texttt{0}&\texttt{0}
\\
\texttt{0}&\lambda^-_j&\texttt{0}&\texttt{0}\\\texttt{0}&\texttt{0}&\lambda^-_j
&\texttt{0}\\\texttt{0}&\texttt{0}
&\texttt{0}&\lambda^-_k\end{matrix}\right),
\quad\quad\quad\quad\quad\quad\tag{\ref{e40}}
\end{align}
\end{subequations}
are the solutions of Eq. \eqref{e10}. Explicitly, together with
Eqs. \eqref{e37}, and \eqref{e39} we have
\begin{equation}\label{e41}
\alpha^\pm_i=-\lambda^\pm_j-\frac{\beta_2}{\beta_3},\quad
\lambda^\pm_k= e^\pm_1+\frac{\beta_2}{\beta_3}-\lambda^\pm_j.
\end{equation}
The domain of validity of these solutions in the parameter space
of the action Eq. \eqref{e1}, and the state space of the ansatz
Eq. \eqref{e8} is governed by the conditions $B^2-4AC>0$, and
$-3(\beta_2)^2/(\beta_3)^2+4\beta_1/\beta_3\geq0$ with the
respective substitutions of $e_2$, and $e_1$ from Eqs.
\eqref{e38}, and \eqref{e39}. The final solution we have to derive
explicitly in this class is $N_9$. If we take its trace we find
that
\begin{equation}\label{e42}
e_1=trN_9=\alpha_i+\lambda_1+\lambda_2=\alpha_i-\frac{B}{A}.
\end{equation}
From this relation by referring to Eq. \eqref{e11} we see that
\begin{equation}\label{e43}
\alpha_i=-\frac{\beta_2}{\beta_3}.
\end{equation}
Also,
\begin{equation}\label{e44}
tr(N_9)^2=(e_1)^2-2e_2=(\alpha_i)^2+(\lambda_1)^2+(\lambda_2)^2.
\end{equation}
By using the identity \eqref{e34} this relation reduces to the
condition
\begin{equation}\label{e45}
(\beta_2)^2-\beta_1\beta_3=0.
\end{equation}
Now, substitution of Eq. \eqref{e43} into the polynomial Eq.
\eqref{e12} gives $e_2$ in terms of $e_1$ explicitly. It reads
\begin{equation}\label{e46}
e_2=-2\frac{\beta_2}{\beta_3}e_1-2\frac{(\beta_2)^2}{(\beta_3)^2}-\frac{\beta_1}{\beta_3}+\frac{D}{\beta_2}.
\end{equation}
We see in this formulation that $e_1$ remains completely an
arbitrary spacetime field. For a particular choice of it one can
read $e_2$ from Eq. \eqref{e46}, and construct $A,B,C$ explicitly
in terms of $\{\beta_i\}$, and $D$ via their definitions in Eq.
\eqref{e11}, then one can explicitly obtain the entries of $N_9$
from
\begin{equation}\label{e47}
\lambda_{1,2}=\frac{-B\pm\sqrt{B^2-4AC}}{2A},
\end{equation}
and Eq. \eqref{e43}. For this solution to exist the conditions Eq.
\eqref{e45}, and $B^2-4AC>0$ must be satisfied. Again, the second
of these defines a domain in the union of the action-parameter
space of the theory, and the state space of the effective fluid.
\subsection{$\Delta=0$ Solutions}
We now turn our attention on the cases when $\Delta=B^2-4AC=0$. In
these cases there is a repeated root $\lambda^\prime=-B/2A$ of the
polynomial Eq. \eqref{e12.1}. The roots of Eq. \eqref{e12.1}
become $\{0,\lambda^\prime,\lambda^\prime\}$. Since, when it is
factorized Eq. \eqref{e14} has a repeated factor some of the
Jordan normal forms which satisfy Eq. \eqref{e14} may have a
repeated root of their minimum polynomials. Thus, when $\Delta=0$
we have nondiagonal, as well as diagonal Jordan normal forms which
satisfy Eq. \eqref{e14}. We again, do not consider the solutions
arising from $B=0$, and/or $C=0$ conditions. Bearing this fact in
mind, therefore, in this case the list of all the possible Jordan
normal forms which satisfy Eq. \eqref{e10} can be given as
\begin{subequations}\label{e48}
\begin{align}
 N_{10}=\left(\begin{matrix}
\alpha_{i}&\texttt{0}&\texttt{0}&\texttt{0}
\\
\texttt{0}&\lambda^\prime
&\texttt{0}&\texttt{0}\\\texttt{0}&\texttt{0}&\lambda^\prime
&\texttt{0}\\\texttt{0}&\texttt{0}
&\texttt{0}&\lambda^\prime\end{matrix}\right),
N_{11}=\left(\begin{matrix}
\alpha_i&\texttt{0}&\texttt{0}&\texttt{0}
\\
\texttt{0}&\lambda^\prime
&\texttt{0}&\texttt{0}\\\texttt{0}&\texttt{0}&\texttt{0}
&\texttt{0}\\\texttt{0}&\texttt{0}
&\texttt{0}&\texttt{0}\end{matrix}\right), N_{12}=
\left(\begin{matrix} \alpha_i&\texttt{0}&\texttt{0}&\texttt{0}
\\
\texttt{0}
&\lambda^\prime&\texttt{0}&\texttt{0}\\\texttt{0}&\texttt{0}&\lambda^\prime
&\texttt{0}\\\texttt{0}&\texttt{0}
&\texttt{0}&\texttt{0}\end{matrix}\right),\notag\\\notag\\
\quad\quad\quad N_{13}=\left(\begin{matrix}
\alpha_{i}&\texttt{0}&\texttt{0}&\texttt{0}
\\
\texttt{0}
&\lambda^\prime&\texttt{1}&\texttt{0}\\\texttt{0}&\texttt{0}&\lambda^\prime
&\texttt{0}\\\texttt{0}&\texttt{0}
&\texttt{0}&\lambda^\prime\end{matrix}\right), \quad
N_{14}=\left(\begin{matrix}
\alpha_i&\texttt{0}&\texttt{0}&\texttt{0}
\\
\texttt{0}
&\lambda^\prime&\texttt{1}&\texttt{0}\\\texttt{0}&\texttt{0}&\lambda^\prime
&\texttt{0}\\\texttt{0}&\texttt{0}
&\texttt{0}&\texttt{0}\end{matrix}\right).
\quad\quad\quad\quad\quad\quad\tag{\ref{e48}}
\end{align}
\end{subequations}
The reader may again verify that these matrices satisfy Eq.
\eqref{e10} by direct substitution. If we take the trace of the
matrices in Eq. \eqref{e48} we find
\begin{equation}\label{e49}
e_1=trN_a=\alpha_i-n\frac{B}{2A},
\end{equation}
where $n=3,1,2,3,2$, for $N_{10},N_{11},N_{12},N_{13},N_{14}$,
respectively. When the definitions in Eq. \eqref{e11} are used
this relation yields
\begin{equation}\label{e50}
\alpha_i=-\frac{n}{2}\frac{\beta_2}{\beta_3}+\frac{2-n}{2}e_1.
\end{equation}
We also have
\begin{equation}\label{e51}
tr(N_a)^2=(e_1)^2-2e_2=(\alpha_i)^2+n(\frac{B}{2A})^2,
\end{equation}
which can be written in the form
\begin{equation}\label{e52}
e_2=-\frac{n^2+n}{8}\frac{(\beta_2)^2}{(\beta_3)^2}-\frac{n^2-n}{4}\frac{\beta_2}{\beta_3}e_1-\frac{n^2-3n}{8}(e_1)^2,
\end{equation}
by using Eqs. \eqref{e11}, and \eqref{e50}. Now, substituting Eqs.
\eqref{e50}, and \eqref{e52} into Eq. \eqref{e12} yields
\begin{equation}\label{e53}
a^\prime(e_1)^3+b^\prime(e_1)^2+c^\prime e_1+d^\prime=0,
\end{equation}
where
\begin{subequations}\label{e54}
\begin{align}
a^\prime &=(-2n+3n^2-n^3)(\beta_3)^3,\notag\\
b^\prime &=(4n+3n^2-3n^3)\beta_2(\beta_3)^2,\notag\\
c^\prime &=(16-8n)\beta_1(\beta_3)^2+(6n-3n^2-3n^3)(\beta_2)^2\beta_3,\notag\\
d^\prime
&=-(3n^2+n^3)(\beta_2)^3-8n\beta_1\beta_2\beta_3+16(\beta_3)^2D.
\tag{\ref{e54}}
\end{align}
\end{subequations}
On the other hand, for these solutions to exist we have the
condition $\Delta=B^2-4AC=0$. Again, by using the definitions in
Eq. \eqref{e11}, and by expressing $e_2$ in terms of $e_1$ via
$\Delta=0$ condition, then by substituting the result in the
equation \eqref{e52} we obtain
\begin{equation}\label{e55}
\frac{n^2-3n+2}{8}(e_1)^2+\frac{n^2-n-2}{4}\frac{\beta_2}{\beta_3}e_1+\frac{n^2+n+2}{8}\frac{(\beta_2)^2}{(\beta_3)^2}-\frac{\beta_1}{\beta_3}=0.
\end{equation}
For $n=1$, this equation is reduced to a linear one and it has the
solution
\begin{equation}\label{e56}
e_1=-\frac{2\beta_1}{\beta_2}+\frac{\beta_2}{\beta_3},
\end{equation}
for $n=2$, the $e_1-$terms vanish and it boils down to the
condition
\begin{equation}\label{e57}
(\beta_2)^2-\beta_1\beta_3=0,
\end{equation}
and for $n=3$, it has the solutions
\begin{equation}\label{e58}
e_1=-\frac{2\beta_2}{\beta_3}\pm
2\sqrt{\frac{\beta_1}{\beta_3}-\frac{3}{4}\frac{(\beta_2)^2}{(\beta_3)^2}},
\end{equation}
provided that $\beta_1/\beta_3-3(\beta_2)^2/4(\beta_3)^2\geq 0$.
When Eqs. \eqref{e56}, and \eqref{e58} are substituted into the
equation \eqref{e53} one finds the equation of state for the
effective fluid. However, at this stage we need not explicitly
derive these expressions since such solutions can not lead to
evolving scale factors for these cases. We will explain the reason
why this occurs for the $n=1$, and $n=3$ cases, and thus, why we
can disregard them in the next section. On the other hand, for the
$n=2$ cases namely for the solutions $N_{12},N_{14}$ Eq.
\eqref{e55} does not fix $e_1$ in terms of the
$\beta_i-$coefficients but only results in a condition on them.
For these cases $e_1$ must be solved from Eq. \eqref{e53} thus,
the equation of state of the effective fluid is not fixed. Solving
Eq. \eqref{e53} which reduces to be a quadratic equation when
$n=2$ for $e_1$ yields
\begin{equation}\label{e59}
e^\pm_1=\bigg[\frac{-c^\prime\pm\sqrt{(c^\prime)^2-4b^\prime
d^\prime}}{2b^\prime}\bigg]_{n=2}.
\end{equation}
Having found $e_1$ now, we can explicitly express the entries of
the solutions $N_{12},N_{14}$ via
\begin{equation}\label{e60}
\lambda^\prime_\pm=-\frac{B}{2A}=\frac{\beta_2}{2\beta_3}+\frac{e^\pm_1}{2},\quad\alpha_i=-\frac{\beta_2}{\beta_3},
\end{equation}
where $e_1$ must be substituted from Eq. \eqref{e59}. We see that,
there are two sets of solutions for each of $N_{12}$, and
$N_{14}$. For the existence of these solutions, there are two
conditions to be satisfied; one of them is Eq. \eqref{e57}, and
the other one is $(c^\prime)^2-4b^\prime d^\prime\geq 0$.
\subsection{$\Delta<0$ solutions}
When $\Delta=B^2-4AC<0$ the polynomial \eqref{e12.1} has complex
roots. The roots of Eq. \eqref{e12.1} are
$\{0,\lambda,\lambda^*\}$. Upon factorization, Eq. \eqref{e14} has
complex root factors too. The non-trivial Jordan normal forms
which satisfy Eq. \eqref{e14} by causing a minimum polynomial that
is a sub-factor in the factorization of Eq. \eqref{e14} to vanish
must have complex eigenvalues. Thus, they are nondiagonal. In this
case, if $\lambda$ is an eigenvalue (or a root of the
corresponding minimum polynomial) of any Jordan normal form which
satisfy Eq. \eqref{e14} $\lambda^*$ must also be an eigenvalue.
Beside this fact, by assuming $B\neq 0$, and $C\neq 0$, also by
considering the form in Eq. \eqref{e13} we conclude that, the only
possible Jordan normal form in this class that would satisfy Eq.
\eqref{e10} is
\begin{equation}\label{e61}
 N_{15}=\left(\begin{matrix}
\alpha_{i}&\texttt{0}&\texttt{0}&\texttt{0}
\\
\texttt{0}&R &I&\texttt{0}\\\texttt{0}&-I&R
&\texttt{0}\\\texttt{0}&\texttt{0}
&\texttt{0}&\texttt{0}\end{matrix}\right),
\end{equation}
where we define $\lambda=R+Ii$ with
\begin{equation}\label{e62}
 R=-\frac{B}{2A},\quad I=\frac{\sqrt{4AC-B^2}}{2A}.
\end{equation}
Again, it can be verified that Eq. \eqref{e61} satisfies Eq.
\eqref{e10} via direct substitution. Now, by taking the trace of
$N_{15}$, and also by referring to Eq. \eqref{e11} we find that
\begin{equation}\label{e63}
\alpha_i=-\frac{\beta_2}{\beta_3}.
\end{equation}
Furthermore,
\begin{equation}\label{e64}
tr(N_{15})^2=(e_1)^2-2e_2=(\alpha_i)^2+2(R^2-I^2),
\end{equation}
which reduces to the condition
\begin{equation}\label{e65}
(\beta_2)^2-\beta_1\beta_3=0,
\end{equation}
upon using the definitions in Eq. \eqref{e11}. Substituting Eq.
\eqref{e63} into Eq. \eqref{e12} leads us to
\begin{equation}\label{e66}
e_2=-\frac{2\beta_2}{\beta_3}e_1-\frac{2(\beta_2)^2}{(\beta_3)^2}-\frac{\beta_1}{\beta_3}+\frac{D}{\beta_2}.
\end{equation}
We realize that, in this solution $e_1$ remains to be an arbitrary
spacetime function. When one specifies $e_1$ one can express $e_2$
in terms of it from Eq. \eqref{e66}, then one can explicitly
obtain the matrix entries of $N_{15}$ by using Eq. \eqref{e11} in
Eq. \eqref{e62}, and from Eq. \eqref{e63}. The conditions of
existence of $N_{15}$ are Eq. \eqref{e65}, and $\Delta=B^2-4AC<0$
in which the particular choice of $e_1$, and the corresponding
$e_2$ must be used.
\section{The solution space}
In the previous section, we have explicitly constructed the entire
set of nontrivial Jordan canonical form solutions of Eq.
\eqref{e10}. We have disregarded the trivial case of
$\sqrt{\Sigma}=\text{diag}(\alpha_i,0,0,0)$ which results in
nonphysical $f-$metric solutions. Before defining the solution
space of Eq. \eqref{e10}, let us discuss one last constraint on
the solutions that we constructed in the previous section. In
obtaining the Jordan normal form of the solutions, although we
used the conditions on the elementary symmetric polynomials $e_1$,
and $e_2$ we did not refer to the $e_3-$structure of the
solutions. This is a necessary, and a crucial point, as our
solution ansatz Eq. \eqref{e8} brings the constraint Eq.
\eqref{e8.6} on $e_1,e_2,e_3$ values of the solutions which we
have not yet considered. To impose this condition on the solutions
we simply take the trace of Eq. \eqref{e10}. Following the trace
operation on Eq. \eqref{e10} if we make use of the Eqs.
\eqref{e3}, and \eqref{e8.6} we get the relation
\begin{equation}\label{e67}
-2\beta_1e_1-\beta_2e_2-\tilde{\rho}+2\tilde{p}=0.
\end{equation}
When, for each solution the appropriate value of
$e_1=e_1(\beta_i,\tilde{\rho},\tilde{p})$, and
$e_2=e_2(\beta_i,\tilde{\rho},\tilde{p})$ of that particular
solution are used in Eq. \eqref{e67} the above expression fixes
the equation of state of the effective fluid that is
$\tilde{p}=\tilde{p}(\tilde{\rho})$ for the solution chosen. At
this stage, we can explain why we have to exclude the solutions of
Eq. \eqref{e10} for the $B=0$, and/or $C=0$ cases. These
conditions via the definitions in Eq \eqref{e11} will bring an
extra constraint on $\tilde{\rho},\tilde{p}$. Thus, when solved
simultaneously with Eq. \eqref{e67} this constraint will cause
$\tilde{\rho}$, and $\tilde{p}$ to be constants. Besides, the
$n=1,3$ cases of the $\Delta=0$ solutions of the previous section
also can be eliminated as cosmological solutions. Similarly, for
those solutions we have seen that the $\Delta=0$ condition already
caused the determination of the equation of state of the effective
fluid composing the cosmological solution ansatz Eq. \eqref{e8}.
For these cases, when one solves the resulting conditions on
$\tilde{\rho},\tilde{p}$ coming from Eqs. \eqref{e53}, and
\eqref{e56}, or \eqref{e58} together with Eq. \eqref{e67} one sees
that both $\tilde{\rho}$, and $\tilde{p}$ must be constants again.
Therefore, as it will be clear in the next section for all of
these cases the scale factor can not evolve hence, it results in
static, and nonphysical cosmological solutions. Now, firstly let
us define the set of Jordan normal form solutions of Eq.
\eqref{e10}
\begin{equation}\label{e68}
\mathcal{J}_1=
\bigg\{N_1,N_2,N_3,N_4,N_5,N_6,N_7,N_8,N_9,N_{12},N_{14},N_{15}\bigg\}.
\end{equation}
Next, we introduce the matrix field
\begin{equation}\label{e69}
P_1={\left(\begin{array}{c|c}  m(x^\mu) &
0    \\
\hline 0& P_{3\times 3}(x^\mu)
\end{array}\right)},
\end{equation}
where $P(x^\mu)$ is an invertible $3\times 3$ matrix field, and
$m(x^\mu)$ is a scalar field which can simply be taken as
$m(x^\mu)=1$ without loss of generality. Since any element
$J\in\mathcal{J}_1$ is a solution of Eq. \eqref{e10} if we perform
a similarity transformation on both sides of Eq. \eqref{e10} we
get
\begin{equation}\label{e70}
A(P_1^{-1}JP_1)^3+B(P_1^{-1}JP_1)^2+CP_1^{-1}JP_1+\mathcal{D}=0,
\end{equation}
where we have used the fact that
$P_1^{-1}\mathcal{D}P_1=\mathcal{D}$. Therefore, we see that
$P_1^{-1}JP_1$ is also a solution of Eq. \eqref{e10} for any
$J\in\mathcal{J}_1$, and for any matrix field of the form Eq.
\eqref{e69}. In that regard, we can define a subset of the
solution space of Eq. \eqref{e10} as
\begin{equation}\label{e71}
\mathcal{M}_1=\bigg\{\sqrt{\Sigma}\mid\sqrt{\Sigma}=P_1^{-1}JP_1\:\:
\big|\:\: \forall J \in \mathcal{J}_1,\: \text{and}\:\:
\text{det}P_1\neq 0\bigg\},
\end{equation}
in which $P_1$ is any matrix field of the form Eq. \eqref{e69}. We
should remark that Eq. \eqref{e67} that is obtained for the
element $J\in\mathcal{J}_1$ remains the same for also the
corresponding element $P_1^{-1}JP_1\in\mathcal{M}_1$ as the
elementary symmetric polynomials do not vary under similarity
transformations. Now, let us consider the matrix equations
\begin{equation}\label{e71.1}
A(J)^3+B(J)^2+CJ+\mathcal{D}_i=0,
\end{equation}
where $i=2,3,4$. Here, $D_2=\text{diag}(0,D,0,0)$,
$D_3=\text{diag}(0,0,D,0)$, and $D_4=\text{diag}(0,0,0,D)$. The
Jordan canonical form solution spaces namely $\mathcal{J}_i$ of
these equations can be constructed from the elements of
$\mathcal{J}_1$. In particular, for example, the diagonal elements
of $\mathcal{J}_2$ are obtained by placing $\alpha_i$ to the
second diagonal entry, and by shifting the rest of the diagonal
entries diagonally in the elements of $\mathcal{J}_1$. Also, the
two nondiagonal elements of $\mathcal{J}_2$ are obtained again, by
placing $\alpha_i$ in the second diagonal entry and by shifting
the primary blocks diagonally in $N_{14},N_{15}$. The elements of
$\mathcal{J}_3$, and $\mathcal{J}_4$ can be obtained in a similar
fashion. We should state that all these diagonal shifting
operations which are used to generate the elements of
$\mathcal{J}_{2,3,4}$ from the elements of $\mathcal{J}_1$ do not
change the elementary symmetric polynomials of the corresponding
element as we keep the diagonal content in these operations. Thus,
the parametrization derived in Section three for the elements of
$\mathcal{J}_1$ are also valid for the elements of
$\mathcal{J}_{2,3,4}$. Next, let us define the invertible $4\times
4$ transformation matrix functions
\begin{equation}\label{e71.2}
 P_2=\left(\begin{matrix}
\texttt{0}&t_2&\texttt{0}&\texttt{0}
\\
\bullet&\texttt{0}&\bullet&\bullet\\\bullet&\texttt{0}&\bullet
&\bullet\\\bullet&\texttt{0} &\bullet&\bullet\end{matrix}\right),
P_3=\left(\begin{matrix}
\texttt{0}&\texttt{0}&t_3&\texttt{0}\\
\bullet&\bullet&\texttt{0}&\bullet\\\bullet&\bullet&\texttt{0}
&\bullet\\\bullet&\bullet &\texttt{0}&\bullet\end{matrix}\right),
P_4= \left(\begin{matrix} \texttt{0}&\texttt{0}&\texttt{0}&t_4
\\
\bullet&\bullet&\bullet&\texttt{0}\\\bullet&\bullet&\bullet
&\texttt{0}\\\bullet&\bullet
&\bullet&\texttt{0}\end{matrix}\right),
\end{equation}
where $t_{2,3,4}$ are arbitrary functions of $x^\mu$ like
$m(x^\mu)$, and the entries which are not specified in the above
$4\times 4$ matrices form up partitioned $3\times 3$ invertible
matrix functions. If now, we define the solution spaces
\begin{subequations}\label{e71.3}
\begin{align}
\mathcal{M}_2&=\bigg\{\sqrt{\Sigma}\mid\sqrt{\Sigma}=P_2JP_2^{-1}\:\:
\big|\:\: \forall J \in \mathcal{J}_2,\: \text{and}\:\:
\text{det}P_2\neq 0\bigg\},\notag\\
\mathcal{M}_3&=\bigg\{\sqrt{\Sigma}\mid\sqrt{\Sigma}=P_3JP_3^{-1}\:\:
\big|\:\: \forall J \in \mathcal{J}_3,\: \text{and}\:\:
\text{det}P_3\neq 0\bigg\},\notag\\
\mathcal{M}_4&=\bigg\{\sqrt{\Sigma}\mid\sqrt{\Sigma}=P_4JP_4^{-1}\:\:
\big|\:\: \forall J \in \mathcal{J}_4,\: \text{and}\:\:
\text{det}P_4\neq 0\bigg\},\notag\\
\tag{\ref{e71.3}}
\end{align}
\end{subequations}
then
\begin{equation}\label{e71.4}
\mathcal{M}=\mathcal{M}_1\cup
\mathcal{M}_2\cup\mathcal{M}_3\cup\mathcal{M}_4,
\end{equation}
becomes the general solution space of the Eq. \eqref{e10}. We will
give a sketch of the proof of this fact in the Appendix. Now that
we have found the complete solution space of Eq. \eqref{e10}, we
can turn our attention on the background metric solutions of the
action Eq. \eqref{e1}. By referring to Eq. \eqref{e3.1} we now
have
\begin{equation}\label{e74}
f=g\Sigma,
\end{equation}
where $\Sigma$ is the square of any element in $\mathcal{M}$, and
$g$ is the FLRW metric. However, not all elements of $\mathcal{M}$
which solve Eq. \eqref{e10} will lead to symmetric results in Eq.
\eqref{e74} thus, physically acceptable background metrics. We
have to impose the condition
\begin{equation}\label{e75}
g\Sigma=\Sigma^Tg,
\end{equation}
which guarantees the symmetry of $f$. Therefore, we define the
cosmological solution moduli of the action Eq. \eqref{e1} as the
set
\begin{equation}\label{e76}
\Gamma_\mathcal{C}=\big\{ (g,f)\:\:\big|\:\: f=gX^2\:\:\big|
X\in\mathcal{M},\:\text{and}\: gX^2=(X^T)^2g\big\}.
\end{equation}
In special, when one chooses the diagonal elements in
$\mathcal{J}_1,\mathcal{J}_2, \mathcal{J}_3,\mathcal{J}_4,$ then
squares them, and substitutes the result in Eq. \eqref{e74} one
directly obtains the exact background metric solutions in a
concise way without being obliged to concern the symmetry
condition. On the other hand, for the more general cases one has
to choose a special form for the matrices $P_1,P_2,P_3,P_4$ to
satisfy Eq. \eqref{e75}. Since, the symmetry requirement in Eq.
\eqref{e75} becomes
\begin{equation}\label{e77}
gP_iJ^2P_i^{-1}=(P_iJ^2P_i^{-1})^Tg,
\end{equation}
a closer inspection denotes that for a particular choice of
$J\in\mathcal{J}_i$ this equation brings three algebraic
constraint conditions on the function-entries of the
solution-generating $P_i-$matrices (in particular, their
unspecified $3\times 3$ partitions) which enable us to determine
three of the entries of these partitions in terms of the other six
entries which remain arbitrary. Next, we will give a summary of
the cosmological dynamics.
\section{Cosmological dynamics}
In the $g-$sector beside the effective ideal fluid energy-momentum
tensor that is introduced in Eq. \eqref{e8}, we will also take the
physical matter as a perfect fluid with the energy-momentum tensor
\begin{equation}\label{e78}
T^{g}_{M\:\mu\nu}=(\rho+p)U_{\mu}U_{\nu}+pg_{\mu\nu},
\end{equation}
where $p=p(t)$, and $\rho=\rho(t)$ are the pressure, and the
energy density of the $g-$matter fluid, respectively. Now, by
using the physical $g-$matter, and the effective energy-momentum
tensors together with the FLRW metric Eq. \eqref{e7.7} in the
$g-$metric equations Eq. \eqref{e4} leads us to the $t-$component
equation
\begin{equation}\label{e79}
\big(\frac{\dot{a}}{a}\big)^2+\frac{k}{a^2}=\frac{8\pi
G}{3}\rho-\frac{m^2}{3}\tilde{\rho}-\frac{\Lambda^g}{6},
\end{equation}
as well as the three identical spatial$-$component equations
\begin{equation}\label{e80}
\frac{2\ddot{a}}{a}=-\big(\frac{\dot{a}}{a}\big)^2-\frac{k}{a^2}-8\pi
Gp+m^2\tilde{p} -\frac{\Lambda^g}{2},
\end{equation}
which become the modified Friedmann equations. By using Eq.
\eqref{e79} in Eq. \eqref{e80} we can obtain the modified cosmic
acceleration equation as
\begin{equation}\label{e81}
\frac{\ddot{a}}{a}=-\frac{4\pi
G}{3}\big(3p+\rho\big)+\frac{m^2}{6}\big(3\tilde{p}+\tilde{\rho}\big)-\frac{\Lambda^g}{6}.
\end{equation}
We observe that the Friedmann, and cosmic acceleration equations
are in the canonical form, only getting additional contributions
from the effective fluid pressure, and energy density which are
the reflections of the interaction Lagrangian term in Eq.
\eqref{e1} which is proportional to the squared graviton mass. The
matter-fluid equation
\begin{equation}\label{e82}
\dot{\rho}=-\frac{3\dot{a}}{a}\big(p+\rho\big),
\end{equation}
is the consequence of the matter energy-momentum conservation law
namely, $\nabla^\mu T^{g}_{M\:\mu\nu}=0$ that is derived for the
FLRW $g-$metric. Besides, a similar continuity equation
\begin{equation}\label{e83}
\dot{\tilde{\rho}}=-\frac{3\dot{a}}{a}\big(\tilde{p}+\tilde{\rho}\big),
\end{equation}
for the effective fluid follows from the substitution of the
effective energy-momentum tensor Eq. \eqref{e8} into the
corresponding Bianchi identity in Eq. \eqref{e7.6} upon using the
FLRW $g-$metric. On the other hand, in this solution scheme the
$f-$metric equation becomes
\begin{equation}\label{e84}
\kappa\big[R^f_{\mu\nu}-\frac{1}{2}R^f
f_{\mu\nu}-\frac{1}{2}\Lambda^f
f_{\mu\nu}\big]+m^2\frac{\sqrt{-g}}{\sqrt{-f}}f_{\mu\rho}(\tilde{\rho}+\tilde{p})\delta^{\rho}_{\:\:0}\delta^{0}_{\:\:\nu}=\epsilon
8\pi GT^{f}_{M\,\mu\nu},
\end{equation}
where we have used Eq. \eqref{e10} in Eq. \eqref{e6}, and
substituted the result in Eq. \eqref{e4.1}. When Eq. \eqref{e83}
is satisfied one does not have to consider the second of the
Bianchi identities in Eq. \eqref{e7.6} as it is also automatically
satisfied \cite{bm2,bm1}. Since the two metric sectors are
efficiently decoupled from each other one can solve the $g-$sector
equations independently without making an assumption on the
$f-$matter. The solution methodology should start by first
choosing which similarity equivalence class representatives in
$\mathcal{J}_{1,2,3,4}$ links the two metrics. Fixing $J$ in this
way determines the equation of state of the effective fluid via
Eq. \eqref{e67} by substituting the appropriate elementary
symmetric polynomials of the chosen $J$. Then, by using the
equations of state of the effective, and the $g-$matter
(corresponding to various eras) one can solve Eqs. \eqref{e79},
\eqref{e82}, and \eqref{e83} to find out the evolution of the
scale factor, and the state of the effective fluid, and the matter
ideal fluid. The reader should appreciate that, our solutions are
justified only if one finds also solutions of the $f-$metric
equations Eq. \eqref{e84}. In general, one can now use
$a,\tilde{\rho},\tilde{p}$ (which are completely determined) in
Eq. \eqref{e74} to read the associated $f-$metric which has an
implicit dynamical link in Eq. \eqref{e74} to the $g-$sector via
barely, the metric $g$, and the elements of
$\mathcal{J}_{1,2,3,4}$ which are not only functions of the
$\beta_i-$parameters of the theory but also the effective
pressure, and the energy density of the effective ideal fluid
whose functional forms are solved from the cosmological equations
of the $g-$metric sector. At this point we have to remark that,
although we have previously mentioned about a degree of
arbitrariness in constructing $f$ in the set Eq. \eqref{e76} via
six arbitrary entries of the matrices $P_i$ these arbitrary
functions may also be used to fix the form of $f$ now, so that it
will satisfy Eq. \eqref{e84} when one chooses the form of
$f-$matter in it. However, this route is not the only one to be
followed in general. On the contrary, to exemplify a solution
outline in the $f-$sector let us consider the solutions $N_1,N_2$.
If they are used in Eq. \eqref{e74} one obtains
\begin{equation}\label{e84.1}
f=-N(t)dt^2+\frac{b^2(t)}{1-kr^2}dr^2+b^2(t)r^2 d\theta^2
+b^2(t)r^2 sin^2 \theta d\varphi^2,
\end{equation}
which is in the generalized FLRW form with a lapse function
$N(t)$. Here, we see that $N(t)=(\alpha_i)^2$, and the $f-$scale
factor can be read from $b^2(t)=(\lambda_{1,2})^2a^2(t)$. We can
also read $\alpha_i=\alpha_i[\beta_j,\tilde{\rho},\tilde{p}]$, and
$\lambda_{1,2}=\lambda_{1,2}[\beta_j,\tilde{\rho},\tilde{p}]$ from
Eq. \eqref{e26}. Since, as we discussed above from the $g-$sector
equations $a,\tilde{\rho},\tilde{p}$ are completely solved $N(t),$
and $b(t)$ in Eq. \eqref{e84.1} are also determined. Thus, in this
case the $f-$metric is fixed as a generalized FLRW one. Let us
also take the $f-$matter in the perfect fluid form, and consider
$\Lambda^f=\Lambda^f(t)$. With these choices, and the substitution
of the $f-$metric from Eq. \eqref{e84.1} the $f-$metric equation
\eqref{e84} becomes
\begin{equation}\label{e84.2}
\kappa\big[G^f_{\mu\nu}-\frac{1}{2}\Lambda^f(t)
f_{\mu\nu}\big]-m^2\tilde{\Lambda}^f(t)N(t)\delta_{\mu
0}\delta_{0\nu}=\epsilon 8\pi
G\big((\rho_f+p_f)U_{\mu}U_{\nu}+p_ff_{\mu\nu}\big),
\end{equation}
where
\begin{equation}\label{e84.3}
\tilde{\Lambda}^f(t)=\sqrt{\frac{1}{\text{det}(N^2_{1,2})}}(\tilde{\rho}+\tilde{p}).
\end{equation}
We should remind the reader that, in Eq. \eqref{e84.2} the only
unknown functions are $\Lambda^f(t),\rho_f(t),p_f(t)$ as the scale
factor $b(t)$ is predetermined. The third term on the left hand
side in Eq. \eqref{e84.2} will only contribute a time-dependent
effective cosmological constant to the $00-$component but not to
the spatial component equations. However, both the $00-$, and the
spatial-component equations will get extra contributions from the
lapse function with respect to the FLRW ones. Therefore, from Eq.
\eqref{e84.2} we will get two modified Friedmann equations which
are algebraic (since the scale factor is already determined) for
the unknown functions $\Lambda^f(t),\rho_f(t),p_f(t)$. There is
also a first-order ordinary differential equation arising from the
fluid equation of the $f-$matter perfect fluid, namely, from
$\nabla^f_\mu(T^{f}_M)^{\mu}_{\:\:\nu}=0$. Finally, from these two
algebraic, and one first-order ordinary differential equations we
can solve the unknown functions $\Lambda^f(t),\rho_f(t),p_f(t)$ to
complete the $f-$sector solution.
\section{Concluding Remarks}
For the massive bigravity theory \cite{hrbg} we constructed the
complete solution moduli space of the $(f,g)$ couples of metrics
which admit a FLRW cosmology in the $g-$sector via the presence of
an effective ideal fluid contribution coming from the interaction
Lagrangian of the mass terms in addition to the matter one. We
employed the cosmological solution ansatz by choosing the
energy-momentum tensor of the interaction terms in the $g-$metric
equations in the form of an effective ideal fluid one. This choice
resulted in a cubic matrix equation for the building block matrix
of the interaction Lagrangian that is composed of the two metrics.
By deriving the general solution space of this nontrivial matrix
equation (whose coefficients are also functions of the elementary
symmetric polynomials of its solutions) we were able to construct
and define the complete solution space of the $(f,g)$ metric
configurations which enable FLRW cosmologies in the $g-$sector
that is modified by an effective ideal fluid whose contributions
are proportional to the square of the graviton mass. Although, we
obtained the general solutions of the ansatz matrix equation we
also discussed that one still has to impose a symmetry condition
on these solutions to construct a symmetric result for the
$f-$metric. Therefore, in spite of the existence of a matrix field
degree of freedom in constructing $f-$solutions out of the
$g-$sector fields one has to render three out of nine function
components of this arbitrary matrix field to satisfy the symmetry
condition we mentioned. Furthermore, we also discussed that one
might also have to fix the remaining degrees of freedom of the
$f-$metric in satisfying the $f-$metric sector field equations in
the presence of $f-$type matter. We have shown that, the
cosmological solution moduli of bigravity that we constructed is
composed of similarity equivalence classes which do not differ
from each other only in their functional form but also in the
equations of state that they impose on the associated effective
ideal fluid they give rise to. Finally, in the last section, we
presented the resulting cosmological equations of the $g$, and the
$f-$metrics for which we shortly discussed the solution flow chart
dictated by the semi-decoupling of the two metric sectors.

The known exact solutions of bigravity can in general be divided
into three groups \cite{8}. There is a class of solutions in which
both metrics are proportional to each other. There exists another
class of spherically symmetric solutions which has a nondiagonal
background metric. There are also solutions including both
diagonal but not proportional $f$, and $g$ metrics. In this work,
we present the complete cosmological background solution space of
the theory. Massive bigravity as a ghost-free massive gravity
theory promises to possess cosmological solutions which can
exhibit late time self-acceleration which could compensate the
dark energy problem in standard cosmology. The background
cosmological solutions
\cite{bac3,com1,com2,bm2,bm1,3,4,6,8,9,10,11,12,yeni1,yeni2}, and
their perturbations and stability issues
\cite{com2,10,11,per1,per2} arising from the above-listed known
solutions have gained a considerable interest and they are under
extensive inspection recently. It has been shown that although
there are stability problems and the perturbations of these
solutions differ from the ones of GR these problems can still be
overcome by turning on the $f-$type matter which is heuristically
interpreted as dark matter \cite{com2,10,11,per1,per2}. We believe
that, apart from its mathematical legitimacy of completeness which
presents an extensive amount of new cosmological solutions of the
theory our derivation of the cosmological background solution
space can also serve for the phenomenology of the theory. We have
found that, in the general similarity equivalence class structure
of the solutions there is a rich variety of functional relations
between the spatial parts of the two metrics unlike the case in
the particular cosmological solution which is widely studied in
the literature. The behavior of the ratio of the $g$, and
$f-$scale factors of this particular solution (which we believe
must be related to the $N_1$, or $ N_2$ solutions we have
discussed) causes early time instabilities of the perturbations
which differ from the GR ones. Therefore, we hope that among the
variety of complete background solutions we have derived there may
exist ones which may admit acceptable perturbation behavior with
respect to GR perturbations. To explicitly construct, and study
the solutions in this direction one may follow two main routes,
one may either inspect the solution behavior in the various
similarity classes one by one or one may attempt to design
particular form of cosmological solutions with or without
$f-$matter which exhibit a stable nature of perturbations within
the solution construction methodology we have discussed. However,
we should also state that in our generally-constructed solution
space, majority of the $f-$metric solutions may fail to exhibit
homogeneity, and/or isotropy behavior. On the other hand, one may
question the necessity of homogeneity, and isotropy in the
$f-$sector since opposite cases may have acceptable results from
the $g-$metric perturbation theory point of view, and in addition
they may lead to interesting variety of dark matter scenarios.
Finally, we point out a possible direction in which one can extend
the results of the present work to study the cosmological
solutions within the newly proposed formalism of ghost-free
effective-metric-matter coupling \cite{yeni4,yeni5,yeni6}.
\section*{Appendix}
In the Appendix, we will give a sketch of the proof of the
statement we made in Section four that any solution of Eq.
\eqref{e10} must belong to the solution set \eqref{e71.4}. First,
let us assume that Eq. \eqref{e10} has a diagonalizable solution
$X_D$ such that
\begin{equation}\label{e90}
\tau(X_{D})+\mathcal{D}=0.
\end{equation}
If we do a similarity transformation which brings $X_D$ to a
diagonal Jordan form $J_D$ on the above equation then we get
\begin{equation}\label{e91}
P^{-1}\tau(X_{D})P+P^{-1}\mathcal{D}P=\tau(J_D)+P^{-1}\mathcal{D}P=0,
\end{equation}
where $J_D=P^{-1}X_DP$. We can directly observe that
$P^{-1}\mathcal{D}P$ must be a diagonal matrix. Furthermore, since
the eigenvalues of $\mathcal{D}$ must be invariant under a
similarity transformation we can conclude that the diagonal matrix
$P^{-1}\mathcal{D}P$ must be in one of the forms;
$\text{diag}(D,0,0,0)$, $\text{diag}(0,D,0,0)$,
$\text{diag}(0,0,D,0)$, or $\text{diag}(0,0,0,D)$. From this
observation we deduce that $P$ must be in one of the forms in Eqs.
\eqref{e69}, or \eqref{e71.2}. This result proves that any
diagonalizable solution of Eq. \eqref{e10} must be an element of
$\mathcal{M}$. On the other hand, let us consider
nondiagonalizable solutions of Eq. \eqref{e10} with real
eigenvalues. They also satisfy
\begin{equation}\label{e92}
\tau(X_{ND})+\mathcal{D}=0,
\end{equation}
which can be brought to a form
\begin{equation}\label{e93}
P^{-1}\tau(X_{ND})P+P^{-1}\mathcal{D}P=\tau(J_{ND})+P^{-1}\mathcal{D}P=0,
\end{equation}
where $J_{ND}=P^{-1}X_{ND}P$ is one of the nondiagonal Jordan
canonical forms
\begin{subequations}\label{e94}
\begin{align}
 J_1&=\left(\begin{matrix}
e&\texttt{1}&\texttt{0}&\texttt{0}
\\
\texttt{0}&e&\texttt{1}&\texttt{0}\\\texttt{0}&\texttt{0}&e
&\texttt{1}\\\texttt{0}&\texttt{0}
&\texttt{0}&e\end{matrix}\right), \quad\:\:\,
J_2=\left(\begin{matrix} e&\texttt{1}&\texttt{0}&\texttt{0}
\\
\texttt{0}&e&\texttt{1}&\texttt{0}\\\texttt{0}&\texttt{0}&e
&\texttt{0}\\\texttt{0}&\texttt{0}
&\texttt{0}&a_4\end{matrix}\right), \quad\:\:\, J_3=
\left(\begin{matrix} a_1&\texttt{0}&\texttt{0}&\texttt{0}
\\
\texttt{0}&e&\texttt{1}&\texttt{0}\\\texttt{0}&\texttt{0}&e
&\texttt{1}\\\texttt{0}&\texttt{0}
&\texttt{0}&e\end{matrix}\right),\notag\\\notag\\
J_4&=\left(\begin{matrix} e&\texttt{1}&\texttt{0}&\texttt{0}
\\
\texttt{0}&e&\texttt{0}&\texttt{0}\\\texttt{0}&\texttt{0}&a_3
&\texttt{0}\\\texttt{0}&\texttt{0}
&\texttt{0}&a_4\end{matrix}\right), \quad J_5=\left(\begin{matrix}
a_1&\texttt{0}&\texttt{0}&\texttt{0}
\\
\texttt{0}&e&\texttt{1}&\texttt{0}\\\texttt{0}&\texttt{0}&e
&\texttt{0}\\\texttt{0}&\texttt{0}
&\texttt{0}&a_4\end{matrix}\right), \quad J_6=
\left(\begin{matrix} a_1&\texttt{0}&\texttt{0}&\texttt{0}
\\
\texttt{0}&a_2&\texttt{0}&\texttt{0}\\\texttt{0}&\texttt{0}&e
&\texttt{1}\\\texttt{0}&\texttt{0}
&\texttt{0}&e\end{matrix}\right),
\quad\quad\quad\quad\quad\quad\tag{\ref{e94}}
\end{align}
\end{subequations}
where in $J_{2,3,4,5,6}$ the distinct diagonal elements
$a_{1,2,3,4}$ may be equal to $e$ or a different value than $e$.
Here, we observe that since $\tau(J_{ND})$ is in uppertriangular
form
\begin{subequations}\label{e95}
\begin{align}
 P^{-1}\mathcal{D}P&=\left(\begin{matrix}
u_1&u_5&u_9&u_{13}
\\
u_2&u_6&u_{10}&u_{14}\\u_3&u_7&u_{11} &u_{15}\\u_4&u_8
&u_{12}&u_{16}\end{matrix}\right)\left(\begin{matrix}
D&\texttt{0}&\texttt{0}&\texttt{0}
\\
\texttt{0}&\texttt{0}&\texttt{0}&\texttt{0}\\\texttt{0}&\texttt{0}&\texttt{0}
&\texttt{0}\\\texttt{0}&\texttt{0}
&\texttt{0}&\texttt{0}\end{matrix}\right)\left(\begin{matrix}
t_1&t_2&t_3&t_{4}
\\
t_5&t_6&t_{7}&t_{8}\\t_9&t_{10}&t_{11} &t_{12}\\t_{13}&t_{14}
&t_{15}&t_{16}\end{matrix}\right)\notag\\\notag\\
&=\left(\begin{matrix} u_1t_1D&u_1t_2D&u_1t_3D&u_{1}t_4D
\\
u_2t_1D&u_2t_2D&u_{2}t_3D&u_{2}t_4D\\u_3t_1D&u_3t_2D&u_{3}t_3D
&u_{3}t_4D\\u_4t_1D&u_4t_2D
&u_{4}t_3D&u_{4}t_4D\end{matrix}\right),
 \quad\quad\quad\quad\quad\quad\tag{\ref{e95}}
\end{align}
\end{subequations}
must be in uppertriangular form too. Thus, its diagonal elements
which are its eigenvalues must be $\{0,0,0,D\}$ as under
similarity transformations eigenvalues and their algebraic
multiplicities are preserved. Therefore, $P^{-1}\mathcal{D}P$ must
be in one of the forms
\begin{subequations}\label{e96}
\begin{align}
\left(\begin{matrix} D&\bullet&\bullet&\bullet
\\
\texttt{0}&\texttt{0}&\bullet&\bullet\\\texttt{0}&\texttt{0}&\texttt{0}
&\bullet\\\texttt{0}&\texttt{0}
&\texttt{0}&\texttt{0}\end{matrix}\right),\quad
\left(\begin{matrix} \texttt{0}&\bullet&\bullet&\bullet
\\
\texttt{0}&D&\bullet&\bullet\\\texttt{0}&\texttt{0}&\texttt{0}
&\bullet\\\texttt{0}&\texttt{0}
&\texttt{0}&\texttt{0}\end{matrix}\right),\quad
\left(\begin{matrix} \texttt{0}&\bullet&\bullet&\bullet
\\
\texttt{0}&\texttt{0}&\bullet&\bullet\\\texttt{0}&\texttt{0}&D
&\bullet\\\texttt{0}&\texttt{0}
&\texttt{0}&\texttt{0}\end{matrix}\right),\quad
 \left(\begin{matrix}
\texttt{0}&\bullet&\bullet&\bullet
\\
\texttt{0}&\texttt{0}&\bullet&\bullet\\\texttt{0}&\texttt{0}&\texttt{0}
&\bullet\\\texttt{0}&\texttt{0} &\texttt{0}&D\end{matrix}\right).
 \tag{\ref{e96}}
\end{align}
\end{subequations}
Let us assume that the nondiagonalizable solution is such that
$P^{-1}\mathcal{D}P$ becomes the second form above. In this case,
since $u_2,$ and $t_2$ can not be zero we see from Eq. \eqref{e95}
that $u_3,u_4,t_1$ must vanish. The corresponding Jordan form must
be $J_6$ because in all the other cases via Eq. \eqref{e93} the
diagonal entries would lead to two inconsistent equations
$Ae^3+Be^2+Ce+D=0$, and $Ae^3+Be^2+Ce=0$\footnote{We refer the
reader to the first footnote in Section three.}. In this
restriction, again from Eq. \eqref{e95} we see that as $t_2$ is
not zero $u_1$ must be zero, also, as $u_2$ is not zero $t_3$ must
be zero. Besides, Eq. \eqref{e95} denotes that when $J_6$ is used
in Eq. \eqref{e93} since $u_2\neq 0$ $t_4$ must be zero.
Furthermore, $P^{-1}P=PP^{-1}=1$ yields
$t_6=t_{10}=t_{14}=u_6=u_{10}=u_{14}=0$. Therefore, we conclude
that in this case $P=P_2$, and
$P^{-1}\mathcal{D}P=\text{diag}(0,D,0,0)$. If we consider the
nondiagonalizable solutions which generate the fourth matrix in
Eq. \eqref{e96} for $P^{-1}\mathcal{D}P$ we realize that since
$u_4,t_4$ are nonzero $t_1,t_2,t_3$ must be zero. Also, similar to
the previous case, since they would result in the inconsistent
equations $Ae^3+Be^2+Ce+D=0$, and $Ae^3+Be^2+Ce=0$ the Jordan
canonical forms $J_1,J_3,J_6$ must be excluded in this case. For
this reason, since $t_4\neq 0$, $u_3$ must be zero as a result of
the substitution of the possible Jordan forms $J_2,J_4,J_5$ in Eq.
\eqref{e93}. In addition, since $\tau(J_2,J_4,J_5)$ can not have
nonzero elements at the fourth column except the fourth row, since
$t_4\neq 0$ via Eq. \eqref{e95} we see that $u_1,u_2$ must be
zero. Upon these substitutions, $P^{-1}P=PP^{-1}=1$ gives
$t_8=t_{12}=t_{16}=u_8=u_{12}=u_{16}=0$. Hence, we observe that
for this case $P=P_4$, and
$P^{-1}\mathcal{D}P=\text{diag}(0,0,0,D)$. Two straightforward,
and similar analysis show also that for the first, and the third
cases in Eq. \eqref{e96} the transformation matrices must be
$P_1$, and $P_3$, also, $P^{-1}\mathcal{D}P=\text{diag}(D,0,0,0)$,
and $P^{-1}\mathcal{D}P=\text{diag}(0,0,D,0)$, respectively.
Besides, the possible Jordan forms for these cases are
$J_{3,5,6}$, and $J_4$, respectively. We observe also that, $J_1$
is not possible for any of the nondiagonalizable solutions with
real eigenvalues. Therefore, this analysis proves that any
nondiagonalizable solution of Eq. \eqref{e10} with real
eigenvalues must be contained in $\mathcal{M}$. Now, let us
consider the nondiagonalizable soltions of Eq. \eqref{e10} with
complex eigenvalues. By applying an appropriate similarity
transformation on Eq. \eqref{e92} we get
\begin{equation}\label{e97}
P^{-1}\tau(X_{ND})P+P^{-1}\mathcal{D}P=\tau(K_{ND})+P^{-1}\mathcal{D}P=0,
\end{equation}
where $K_{ND}=P^{-1}X_{ND}P$ is one of the nondiagonal Jordan
canonical forms
\begin{subequations}\label{e98}
\begin{align}
 K_1&=\left(\begin{matrix}
R_1&I_1&\texttt{0}&\texttt{0}
\\
-I_1&R_1&\texttt{0}&\texttt{0}\\\texttt{0}&\texttt{0}&R_2
&I_2\\\texttt{0}&\texttt{0}
&-I_2&R_2\end{matrix}\right),\quad\quad\: K_2=\left(\begin{matrix}
R&I&\texttt{0}&\texttt{0}
\\
-I&R&\texttt{0}&\texttt{0}\\\texttt{0}&\texttt{0}&e&
\texttt{1}&\\\texttt{0}&\texttt{0}
&\texttt{0}&e\end{matrix}\right),\notag\\\notag\\K_3&=
\left(\begin{matrix} R&I&\texttt{0}&\texttt{0}
\\
-I&R&\texttt{0}&\texttt{0}\\\texttt{0}&\texttt{0}&a_3&
\texttt{0}&\\\texttt{0}&\texttt{0}
&\texttt{0}&a_4\end{matrix}\right),\quad\quad \quad
K_4=\left(\begin{matrix} a_1&\texttt{0}&\texttt{0}&\texttt{0}
\\
\texttt{0}&R&I&\texttt{0}\\\texttt{0}&-I&R
&\texttt{0}\\\texttt{0}&\texttt{0}
&\texttt{0}&a_4\end{matrix}\right),\notag\\\notag\\
K_5&=\left(\begin{matrix} e&\texttt{1}&\texttt{0}&\texttt{0}
\\
\texttt{0}&e&\texttt{0}&\texttt{0}\\\texttt{0}&\texttt{0}&R
&I\\\texttt{0}&\texttt{0} &-I&R\end{matrix}\right),\quad\quad
\quad \quad \: \: K_6= \left(\begin{matrix}
a_1&\texttt{0}&\texttt{0}&\texttt{0}
\\
\texttt{0}&a_2&\texttt{0}&\texttt{0}\\\texttt{0}&\texttt{0}&R
&I\\\texttt{0}&\texttt{0} &-I&R\end{matrix}\right).
\quad\quad\quad\quad\quad\quad\tag{\ref{e98}}
\end{align}
\end{subequations}
We should state that, when substituted into $\tau$ all these
matrices keep their forms with entries changed. For example,
\begin{equation}\label{e99}
\tau(K_{1})=\left(\begin{matrix} R'_1&I'_1&\texttt{0}&\texttt{0}
\\
-I'_1&R'_1&\texttt{0}&\texttt{0}\\\texttt{0}&\texttt{0}&R'_2
&I'_2\\\texttt{0}&\texttt{0} &-I'_2&R'_2\end{matrix}\right).
\end{equation}
If we use this in Eq. \eqref{e97}, and refer to Eq. \eqref{e95} we
see that $u_1t_1=u_2t_2$, and $u_2t_1=-u_1t_2$ which give
$t_1=t_2=0$. Also, $u_3t_3=u_4t_4$, and $u_4t_3=-u_3t_4$ which
give $t_3=t_4=0$. However, now $P$ becomes a zero-matrix, hence,
it becomes singular, and can not perform any similarity
transformation. Therefore, this case must be excluded (there
exists no nonsingular matrix which can bring a nondiagonalizable
solution of Eq. \eqref{e10} into $K_1-$form since, this results in
an inconsistency). Let us consider the case $K_4$ which leads to
the form
\begin{equation}\label{e100}
\tau(K_{4})=\left(\begin{matrix}
a'_1&\texttt{0}&\texttt{0}&\texttt{0}
\\
\texttt{0}&R'&I'&\texttt{0}\\\texttt{0}&-I'&R'
&\texttt{0}\\\texttt{0}&\texttt{0}
&\texttt{0}&a'_4\end{matrix}\right).
\end{equation}
Similarly, now, from Eq. \eqref{e97} we have $u_2t_2=u_3t_3$, and
$u_2t_3=-u_3t_2$ which give $t_3=t_2=0$. From Eq. \eqref{e95} we
deduce that, for a consistent nontrivial $P$ we can not have
$t_1=0$, and $t_4=0$ at the same time. Also, $u_{1,2,3,4}$ can not
vanish simultaneously. These facts leave us two cases. Either;
$t_1=0,t_4\neq 0$, but $u_{1,2,3}=0, u_4\neq 0$, or $t_1\neq
0,t_4= 0$, but $u_{2,3,4}=0, u_1\neq 0$ to have consistency when
Eq. \eqref{e100} is substituted into Eq. \eqref{e97}. The first
case gives $P=P_4$, and $P^{-1}\mathcal{D}P=\text{diag}(0,0,0,D)$
(via the preservation of the eigenvalues under similarity
transformations). Whereas, the second case corresponds to $P=P_1$,
and $P^{-1}\mathcal{D}P=\text{diag}(D,0,0,0)$. Next, let us
consider $K_2$. Similar to the previous cases above now, for the
consistency of Eq. \eqref{e97} we must have $t_1=t_2=0$. Hence,
$P^{-1}\mathcal{D}P$ must be in uppertriangular form with diagonal
elements as its eigenvalues which must be the set $\{D,0,0,0\}$
where $D$ must be either at the third, or the fourth diagonal
entry. However, for either of these cases Eq. \eqref{e97} leads us
to two inconsistent equations $Ae^3+Be^2+Ce+D=0$, and
$Ae^3+Be^2+Ce=0$ as we assumed $D\neq 0$. Therefore, this case
must be excluded. $K_3$ on the other hand, leads to the conditions
$u_1t_1=u_2t_2$, and $u_2t_1=-u_1t_2$ that give $t_1=t_2=0$.
Again, Eq. \eqref{e95} shows that for a consistent nontrivial $P$
we can not have $t_3=0$, and $t_4=0$ at the same time, as well as
$u_{1,2,3,4}$ can not vanish all. Thus, either; $t_3=0,t_4\neq 0$,
but $u_{1,2,3}=0, u_4\neq 0$, or $t_3\neq 0,t_4= 0$, but
$u_{1,2,4}=0, u_3\neq 0$ to have consistency when $\tau(K_3)$ is
used in Eq. \eqref{e97}. The first case gives $P=P_4$, and
$P^{-1}\mathcal{D}P=\text{diag}(0,0,0,D)$, and the second case
corresponds to $P=P_3$, and
$P^{-1}\mathcal{D}P=\text{diag}(0,0,D,0)$. A very similar line of
reasoning denotes that $K_5$ is not possible, also, $K_6$ is
possible with either; $P=P_1$, and
$P^{-1}\mathcal{D}P=\text{diag}(D,0,0,0)$, or $P=P_2$, and
$P^{-1}\mathcal{D}P=\text{diag}(0,D,0,0)$. Therefore, we conclude
that any nondiagonalizable solution with complex eigenvalues of
Eq. \eqref{e10} must also be contained in $\mathcal{M}$. As a
final remark, in summary, in the Appendix we showed that any
diagonalizable or nondiagonalizable solution of Eq. \eqref{e10}
must be an element of $\mathcal{M}$ via proving that their Jordan
canonical forms must satisfy one of the four equations in Eq.
\eqref{e10}, and Eq. \eqref{e71.1}.
\section*{Acknowledgements}
We thank Merete Lillemark for useful communications.


\begin{thebibliography}{99}
\bibitem{dgrt1}
  de Rham C., and Gabadadze G.
  ``\textit{Generalization of the Fierz-Pauli Action}'',
   2010 \textit{Phys. Rev.}  \textbf{D82} 044020
  arXiv:1007.0443 [hep-th].
\bibitem{dgrt2}
  de Rham C., Gabadadze G., and Tolley A. J.
  ``\textit{Resummation of Massive Gravity}'',
  2011  \textit{Phys. Rev. Lett.}  \textbf{106} 231101
  arXiv:1011.1232 [hep-th].
  \bibitem{fp}
  Fierz M., and Pauli W.
  \textit{``On Relativistic Wave Equations for Particles of Arbitrary Spin in an Electromagnetic
  Field}'',
1939   \textit{Proc. Roy. Soc. Lond.} \textbf{A173} 211.
\bibitem{BD1}
  Boulware D. G., and Deser S.
  \textit{``Can Gravitation have a Finite Range?}'',
1972  \textit{Phys. Rev.} \textbf{D6} 3368.
\bibitem{BD2}
  Boulware D. G., and Deser S.
  ``\textit{Inconsistency of Finite Range Gravitation}'',
1972  \textit{Phys. Lett.} \textbf{B40} 227.
\bibitem{hr1}
  Hassan S. F., and Rosen R. A.
  ``\textit{On Non-Linear Actions for Massive Gravity}'',
  2011 \textit{JHEP} \textbf{1107} 009
  arXiv:1103.6055 [hep-th].
\bibitem{hr2}
  Hassan S. F., and Rosen R. A.
  ``\textit{Resolving the Ghost Problem in non-Linear Massive
  Gravity}'',
2012   \textit{Phys. Rev. Lett.} \textbf{108} 041101
  arXiv:1106.3344 [hep-th].
\bibitem{hr3}
  Hassan S. F., Rosen R. A., and Schmidt-May A.
  ``\textit{Ghost-free Massive Gravity with a General Reference
  Metric}'',
 2012  \textit{JHEP} \textbf{1202} 026
  arXiv:1109.3230 [hep-th].
\bibitem{hrbg}
Hassan S. F., and Rosen R. A.
  ``\textit{Bimetric Gravity from Ghost-free Massive Gravity}'',
  2012 \textit{JHEP} \textbf{1202} 126
  arXiv:1109.3515 [hep-th].
  \bibitem{bac1}
Baccetti V., Martin-Moruno P., and Visser M.
  ``\textit{Massive gravity from bimetric gravity}'',
 2013 \textit{Class. Quant. Grav.} \textbf{30} 015004
arXiv:1205.2158 [gr-qc].
\bibitem{bac2}
Baccetti V., Martin-Moruno P., and Visser M.
  \textit{``Null Energy Condition violations in bimetric gravity''}, 2012
  \textit{JHEP} \textbf{1208} 148
  arXiv:1206.3814 [gr-qc].
  \bibitem{bac3}
Baccetti V., Martin-Moruno P., and Visser M.
\textit{``Gordon and
Kerr-Schild ansatze in massive and bimetric
  gravity''}, 2012
  \textit{JHEP} \textbf{1208} 108
  arXiv:1206.4720 [gr-qc].
\bibitem{com1}
Comelli D., Crisostomi M., Nesti F., and Pilo L.
  \textit{``FRW Cosmology in Ghost Free Massive Gravity''}, 2012
  \textit{JHEP} \textbf{1203} 067
  [Erratum-ibid.\  \textbf{1206} 020 (2012)]
  arXiv:1111.1983 [hep-th].
\bibitem{com2}
Comelli D., Crisostomi M., and Pilo L.
  \textit{``Perturbations in Massive Gravity Cosmology''}, 2012
  \textit{JHEP} \textbf{1206} 085
  arXiv:1202.1986 [hep-th].
\bibitem{bm2}
von Strauss M., Schmidt-May A., Enander J., Mortsell E., and
Hassan S. F. \textit{``Cosmological Solutions in Bimetric Gravity
and their Observational Tests''}, 2012 \textit{JCAP} \textbf{1203}
042 arXiv:1111.1655 [gr-qc].
\bibitem{bm1}
Volkov M. S. \textit{``Cosmological Solutions with Massive
Gravitons in the Bigravity Theory''}, 2012 \textit{JHEP}
\textbf{1201} 035 arXiv:1110.6153 [hep-th].
\bibitem{3}
Volkov M. S. \textit{``Exact Self-Accelerating Cosmologies in the
Ghost-Free Bigravity and Massive Gravity''}, 2012 \textit{Phys.
Rev.} \textbf{D86} 061502 arXiv:1205.5713 [hep-th].
\bibitem{4}
Akrami Y., Koivisto T. S., and Sandstad M.
  \textit{``Accelerated Expansion from Ghost-Free Bigravity: A Statistical Analysis with Improved
  Generality''}, 2013
  \textit{JHEP} \textbf{1303} 099
  arXiv:1209.0457 [astro-ph.CO].
\bibitem{6}
Volkov M. S.
  \textit{``Hairy Black Holes in the Ghost-Free Bigravity Theory''},
  2012
  \textit{Phys. Rev.} \textbf{D85} 124043
  arXiv:1202.6682 [hep-th].
  \bibitem{8}
Volkov M. S.
  \textit{``Self-Accelerating Cosmologies and Hairy Black Holes in Ghost-Free Bigravity and Massive Gravity''},
2013 \textit{Class. Quant. Grav.} \textbf{30} 184009
  arXiv:1304.0238 [hep-th].
\bibitem{9}
Koennig F., Patil A., and Amendola L. \textit{``Viable
Cosmological Solutions in Massive Bimetric Gravity''}, 2014
  \textit{JCAP} \textbf{1403} 029
  arXiv:1312.3208 [astro-ph.CO].
\bibitem{10}
De Felice A., G\"{u}mr\"{u}k\c{c}\"{u}o\u{g}lu A. E., Mukohyama
S., Tanahashi N., and Tanaka T.
  \textit{``Viable Cosmology in Bimetric Theory''}, 2014
  \textit{JCAP} \textbf{1406} 037
  arXiv:1404.0008 [hep-th].
\bibitem{11}
Koennig F., Akrami Y., Amendola L., Motta M., and Solomon A. R.
   \textit{``Stable and Unstable Cosmological Models in Bimetric Massive Gravity''}, 2014
  \textit{Phys. Rev.} \textbf{D90}, no.12, 124014
  arXiv:1407.4331 [astro-ph.CO].
 \bibitem{12}
 Hassan S. F., Schmidt-May A., and von Strauss M.
  \textit{``Particular Solutions in Bimetric Theory and their
  Implications''}, 2014
  \textit{Int. J. Mod. Phys.} \textbf{D23}, no.13, 1443002
  arXiv:1407.2772 [hep-th].
\bibitem{yeni1}
Katsuragawa T.
  \textit{``Properties of Bigravity Solutions in a Solvable
  Class''}, 2014
  \textit{Phys. Rev.}\ \textbf{D89}, no.12, 124007
  arXiv:1312.1550 [hep-th].
  \bibitem{yeni2}
Fasiello M., and Tolley A. J.
   \textit{``Cosmological Stability Bound in Massive Gravity and
   Bigravity''}, 2013
  \textit{JCAP} \textbf{1312}, 002
  arXiv:1308.1647 [hep-th].
  \bibitem{decoup}
  Y$\i$lmaz N. T. \textit{``Decoupling Solution Moduli of Bigravity''}, arXiv:1502.00463 [hep-th].
\bibitem{massgrav}
  Y$\i$lmaz N. T.
  \textit{``Effective Matter Cosmologies of Massive Gravity I: non-Physical
Fluids''}, 2014 \textit{JCAP} \textbf{1408} 037
  arXiv:1405.6402 [hep-th].
  \bibitem{mgrphysfluid}
Y$\i$lmaz N. T.
  \textit{``Effective Matter Cosmologies of Massive Gravity: Physical
  Fluids''}, 2014
  \textit{Phys. Rev.} \textbf{D90}, no.12 124034
arXiv:1412.4919 [hep-th].
\bibitem{per1}
Comelli D., Crisostomi M., and Pilo L.
   \textit{``FRW Cosmological Perturbations in Massive
   Bigravity''}, 2014
  \textit{Phys. Rev.} \textbf{D90}, no.8, 084003
  arXiv:1403.5679 [hep-th].
\bibitem{per2}
De Felice A., Nakamura T., and Tanaka T.
  \textit{``Possible Existence of Viable Models of Bi-Gravity with Detectable Graviton Oscillations by Gravitational Wave Detectors''},
  2014 \textit{PTEP} \textbf{2014}, no.4, 043E01
  arXiv:1304.3920 [gr-qc].
\bibitem{yeni4}
de Rham C., Heisenberg L., and Ribeiro R. H.
  \textit{``On Couplings to Matter in Massive (Bi-)Gravity''},
  2015
  \textit{Class. Quant. Grav.} \textbf{32}, no.3, 035022
  arXiv:1408.1678 [hep-th].
\bibitem{yeni5}
  G\"{u}mr\"{u}k\c{c}\"{u}o\u{g}lu A. E., Heisenberg L., and Mukohyama S.
   \textit{``Cosmological Perturbations in Massive Gravity with Doubly Coupled
   Matter''}, 2015
  \textit{JCAP} \textbf{1502}, no.02, 022
  arXiv:1409.7260 [hep-th].
  \bibitem{yeni6}
Hassan S. F., Kocic M., and Schmidt-May A.
  \textit{``Absence of Ghost in a New Bimetric-Matter Coupling''},
  arXiv:1409.1909 [hep-th].
\end{thebibliography}
\end{document}